\documentclass[twocolumn,twocolappendix]{aastex63}
\usepackage{amsmath}
\usepackage{amssymb}
\usepackage{bm}
\usepackage{graphicx}
\usepackage{color}
\usepackage{float}

\definecolor{darkgreen}{rgb}{0,0.35,0}
\definecolor{blue}{rgb}{0,0,1}
\usepackage{CJKutf8}            % Chinese name

\newcommand{\be}{\begin{eqnarray}}
\newcommand{\ee}{\end{eqnarray}}

\renewcommand{\vec}[1]{{\bm #1}}

\newcommand\as{\bgroup\markoverwith{\textcolor[rgb]{.5, 0, .6}{\rule[0.5ex]{8pt}{1.5pt}}}\ULon}

\newcommand\ys{\bgroup\markoverwith{\textcolor[rgb]{1.0, 0, 1.0}{\rule[0.5ex]{8pt}{1.5pt}}}\ULon}

\defcitealias{Li2021}{L21}

\shorttitle{Migration of Accreting Planets}
\shortauthors{Li et al.}

\begin{document}

\title{Concurrent Accretion and Migration of Giant Planets in their Natal Disks with Consistent Accretion Torque}

%\maketitle

\author[0000-0002-7329-9344]{Ya-Ping Li\begin{CJK*}{UTF8}{gbsn}(李亚平)\end{CJK*}}
%\affiliation{Theoretical Division, Los Alamos National Laboratory, Los Alamos, NM 87545, USA}
\affiliation{Key Laboratory for Research in Galaxies and Cosmology, Shanghai Astronomical Observatory, Chinese Academy of Sciences, Shanghai 200030, China}
\author[0000-0003-3792-2888]{Yi-Xian Chen\begin{CJK*}{UTF8}{gbsn}(陈逸贤)\end{CJK*}}
\affiliation{Department of Astrophysics, Princeton University, Princeton, NJ 08544, USA} 
\author[0000-0001-5466-4628]{Douglas N. C. Lin\begin{CJK*}{UTF8}{gbsn}(林潮)\end{CJK*}}
%[0000-0001-5466-4628]
\affiliation{Department of Astronomy \& Astrophysics, University of California, Santa Cruz, CA 95064, USA}
\affiliation{Institute for Advanced Studies, Tsinghua University, Beijing 100086, China}

\correspondingauthor{Ya-Ping Li}
\email{liyp@shao.ac.cn}

\begin{abstract}
Migration commonly occurs during the epoch of planet formation.  For emerging 
gas giant planets, it proceeds concurrently with their growth through the 
accretion of gas from their natal protoplanetary disks.
Similar migration process should also be applied to the stellar-mass black holes embedded in active galactic nucleus disks. In this work, we perform high resolution 3D and 2D numerical hydrodynamical 
simulations to study the migration dynamics for accreting embedded objects 
over the disk viscous timescales in a self-consistent manner. 
We find that an accreting planet embedded in a predominantly viscous disk has a tendency to migrate outward, in contrast to the inward orbital 
decay of non-accreting planets. 
3D and 2D simulations find the consistent outward migration results for the accreting planets. Under this circumstance, the accreting planet's outward migration 
is mainly due to the asymmetric spiral arms feeding from the global disk into the Hill radius. This is analogous to the unsaturated corotation torque although the imbalance is due to material accretion
within the libration timescale rather than diffusion onto the inner disk. In a disk with a relatively small viscosity, the accreting planets clear deep gaps near their orbits. 
The tendency of inward migration is recovered, albeit with suppressed rates. 
By performing a parameter survey with a range of disks' viscosity, we find that the transition from outward to inward migration occurs with the effective viscous efficiency factor $\alpha\sim 0.003$ for Jupiter-mass planets.
\end{abstract}
\keywords{Accretion (14), Protoplanetary disks (1300), Extrasolar gaseous giant planets (509), Tidal interaction (1699), Black holes (162), Hydrodynamical simulations (767)}

\section{Introduction}\label{sec:intro}

Several distinct exoplanet populations have been detected so far, among which super-Earths, hot Jupiters, and cold Jupiters are three major categories \citep[e.g., see recent review by][]{ZhuDong2021}.
The close-in orbits of super-Earths \citep{Borucki2011} and hot Jupiters \citep{Mayor1995} have motivated the theoretical
hypothesis \citep{Lin1996} that planets may migrate extensively during their
formation in their natal disks \citep{GT1980, Lin1986}. 
The discovery of multiple planets in/near mean motion resonant configurations provides another compelling evidence for disk-driven migration \citep[e.g., see recent review by][]{Weiss2023}. 
This migration arises from the density waves excited by the planets near the Lindblad and corotation resonances \citep{GT1980} or horseshoe drag \citep{Ward1991,Masset2001}.

With a low-mass planet embedded in a relatively viscous protoplanetary disk (PPD), for which the gravitational perturbation on the disk is weak, this corresponds to the regime of the classical type I migration.
The dissipation of the density waves leads to a net torque on the planet, 
which is usually negative driving the planet migrating inward
\citep[e.g., see][for reviews]{PapaloizouLin1984,LinPapaloizou1993, 
KleyNelson2012,Paardekooper2023}. 
The migration timescale associated with the type I torque is usually much shorter than the gas accretion time onto the core, 
posing a serious challenge for the planet formation theory. 
When the planet mass become comparable to disk thermal mass (usually around Jupiter mass), the density waves can
deposit strong torques in the disk through shocks, carving out a low density gap along its orbit. 
This leads
to the type II migration \citep{LinPapaloizou1986a}. 
In the classical type II migration, the gap is so deep that the flow crossing the gap is severely quenched \citep{Bryden2000}.
In this limit, the planet motion would be locked to the viscous drift velocity of the background disk.

However, recent hydrodynamical simulations show that the gap opened by the planet is not fully depleted, which allows the gas flow across the gap to maintain a steady gas profile \citep{Duffell2014,Fung2014,DurmannKley2015,Robert2018,Chen2020b,Li2023}.
Consequently, for example, one new paradigm for the type II migration suggests that the migration rate can be continuously transited from the type I migration formula after considering the steady gap structure \citep{Kanagawa2018}. 

Although the type II migration is generally slower than that of the type I migration, 
it still poses a challenge to the retention of cold Jupiters when compared with 
the observed period distribution \citep{Nelson2000,Ida2013}. 
Some scenarios
have been investigated to identify processes which may lead to 
the slowing down of migration speed or even the reversal of migration direction. 
They include super-Jupiter planets \citep{Dempsey2021}, 
planets located near a cavity \citep[e.g.,][]{Masset2006,Hu2016,Liu2017}, 
in disks with steep surface density profiles \citep{Chen2020b} or magnetized 
disk winds \citep{Aoyama2023}, in radiative and/or adiabatic disks \citep[e.g.,]
[]{Masset2009,Paardekooper2010,Lega2014}.

In the widely adopted core accretion model for giant planet formation, the dynamical 
gas accretion inevitably occurs after the atmosphere mass grows beyond the core mass 
\citep{Bodenheimer1986,Pollack1996,Ida2004,Chen2020,Zhong2022}.  After the embedded
protoplanets have acquired sufficient mass to open gaps in their natal disks, 
the presence of residual gas near their orbits naturally preserves the prospect
of concurrent gas accretion.  This process is another potential factor which could 
impose a large impact on the planet migration dynamics. 

Some  hydrodynamical simulations on the accretion dynamics for the giant planets (or 
companions) have been performed over the last decade \citep[e.g.,]
[]{Kley2001,DAngelo2003,DAngelo2008,Machida2010,Bodenheimer2013,Choksi2023,Li2021,Li2023}, 
there have been very little discussion on its role in migration dynamics. For example, 
\citet{DAngelo2003,DAngelo2008} carried out simulation with a nested grid to achieve high resolutions around the planet,
but they adopted different outer boundary conditions that cannot maintain the desired mass accretion and angular momentum flux in the outer disk (see Section~\ref{sec:bd}).
This can lead to unrealistic inner and outer disk structures, and thus fail to model the coupled migration in a self-consistent way. Another effect is that those earlier simulations
usually did not run long enough to reach the viscous steady state for a relatively low viscosity.

There have been several analytical attempts and 
N-body simulations that incorporate the planet accretion effect on the 
migration dynamics. These approximations are based on the (non
self-consistent) fitting formula of planetary accretion rates from 
prior and separate hydrodynamical simulations \citep{TanigawaTanaka2016, 
Tanaka2020, Wang2020, Jiang2023}. \citet{Robert2018} used a prescribed planetary 
accretion rate to study its effect on the migration.  They found 
the planet migration is further suppressed by the gas depletion due 
to the rapid gas accretion by the planet.  However, realistic accretion 
by giant planets is expected to quench diffusion across the gap and
deplete the inner disk region on a viscous timescale \citep{Li2023}, when the accretion rate approaches the disk supply.
This effect is expected to strengthen the net inward Lindblad torque 
in a way that is not captured by their study.

The observational implications of accreting planets have been studied using  
smoothed particle hydrodynamics (SPH) simulations by \citet{Toci2020}. However,
their simulations did not run long enough to cover the disk viscous time, and 
the dynamical effect on the planet migration has not been fully explored.
\citet{DurmannKley2017} investigated the accretion rate of migrating planets 
with a very large sink hole radius ($r_{\rm a}=0.5 R_{\rm H}$, where $R_{\rm H}$ is the Hill radius). This prescription 
introduces fluctuations in the planet's accretion rates which are difficult 
to converge with accretion parameters. 

To our best knowledge, a fully self-consistent detailed treatment 
for an accreting planet coupled with the disk evolution is still 
lacking in almost all previous investigations on the migration dynamics 
using hydrodynamical simulations.  The disk background determines 
the accretion rate onto the planet, and the gas accretion could in 
turn modify the global disk structure, especially the gap structures 
around the planet, which then alters the migration torque on the planet. 
In this sense, the planet migration should be closely coupled with 
planetary accretion dynamics.

The hydrodynamical evolution of accreting binary in a circumbinary disk has been 
studied for different mass ratios.  It has been found that the secondary has a 
tendency to migrate outward for the mass ratio $q \gtrsim0.05$ and to migrate 
inward for a smaller mass ratio (i.e., $0.01\lesssim q\lesssim0.05$; e.g., 
\citealt{Duffell2020b,Lai2023}). However, the mass ratio for the circumbinary disk 
simulations (e.g., $q\gtrsim0.01$) is usually much larger than the planet-star case, 
it is, therefore, still unclear how do planets migrate when they are actively accreting 
from the embedded disks.
This physical process could be also relevant to the dynamics of the accreting stellar-mass black holes (BHs)/compact objects embedded in active galactic nucleus (AGN) disks. 
The merger of these compact objects in AGN disks is one important channel for gravitational waves.

In this work, we carry out 3D/2D high resolution, global hydrodynamical simulations to study the evolution of planetary migration and accretion 
in a self-consistent way. 
Based on the insights draw from circumbinary disk 
simulations, 
we consider not only the gravitational but also the accretion 
component of the migration torque.
The paper is organized as follows. 
We describe the method and outline 
the model setup for our simulations in Section~\ref{sec:method}.  
Computational results are analyzed in Section~\ref{sec:results}. 
Summary and discussions are presented in the Section~\ref{sec:conc}.

\section{Method}\label{sec:method}

We simulate globally the accreting planet embedded in a thin and non-self-gravitating disk with \texttt{Athena++}
\citep{Stone2020}. 
Our simulations can also be applied to an accreting stellar-mass BH embedded in AGN disks. From now on, we will use the term ``planet" to refer to the embedded object.

High-resolution 3D and 2D simulations are carried out to study 
the effect of accretion on the planet migration. 
We numerically solve the continuity equation and the equation of motion in a frame corotating with the planet.
The origin of our coordinate system is located at the position of the central star with 
mass $M_{*}$. 

We adopt a simple model in which the PPD's temperature is independent of the distance above the mid-plane $T_{\rm disk}\propto r^{-1.0}$ with an aspect ratio of $h_{0}\equiv H/r=0.05$ at $r_{0}$, where $r_{0}$ is a typical scale length of the disk. 
For our 3D models,  we use spherical coordinates $(r, \theta, \phi)$ for the simulations,
and initialize density structure in the cylindrical radial ($R= r \sin \theta$) and vertical ($z = r \cos \theta$) direction as \citep{Nelson2013} \footnote{There is a typo for the radial dependence of density profile presented in Equation~1 of \citet{Li2023}. It should be the cylindrical radius $R$ not spherical $r$ in that equation.}

\begin{equation}
\rho = \rho_{0}\left(\frac{R}{r_{0}}\right)^{p}\exp\left[\frac{GM_{*}}{c_{s}^2} \left(\frac{1}{\sqrt{R^2+z^2}}-\frac{1}{R}\right) \right], 
\end{equation}
and the rotational velocity is

\begin{equation}
v_{\phi}(R,z) = v_{\rm K}\left[ (p+\zeta)\left(\frac{c_{s}}{v_{\rm K}}\right)^{2}+1+\zeta-\frac{\zeta R}{\sqrt{R^2+z^2}}\right]^{1/2}, 
\end{equation}
where $p=-1.5$ is the radial power-law index for gas density, $\zeta=-1.0$ the temperature power-law index, $v_{\rm K}=\sqrt{GM_{*}/R}$.

In calculating the gravitational potential of the planet at $\vec r$, we use a smoothed potential 
of the form \citep[e.g.,][]{GT1980}

\begin{equation}
\Phi_{\rm p}=-\frac{G M_{\mathrm{p}}}{(\left|\vec{r}_{\rm p}-\vec{r}\right|^2+\epsilon^2)^{1/2}}+q \Omega_{\rm p}^{2} \vec r_{\rm p} \cdot \vec r,
\label{potential}
\end{equation}
where $\vec{r}_{\rm p}$ indicates the location of the planet, $\epsilon$ is the softening length and $\epsilon=0.1\ R_{\rm H}$ is adopted unless otherwise stated, $q=M_{\rm p}/M_{*}$ is the mass ratio between the planet and the host star. 
In this study, we consider a planet in a 
fixed circular orbit considered with $r_{\rm p}=a_{\rm p}$ where $a_{\rm p}$ is the 
planet's orbital semi major axis.  The angular velocity of the planet $\Omega_{\rm p}
= (G M_\ast/a_{\rm p}^3)^{1/2}$.   The second term on the 
right hand side of the above equation corresponds to the indirect term associated with the host star's motion relative to 
the center of mass.  
Without the loss of generality, we can consider a solar-type host star with $M_{*}=M_{\odot}$ and therefore $q=0.001$ corresponds to a planet mass of $M_{\rm p}=M_{\rm J}$.

We adopt the conventional $\alpha$ prescription for 
the kinematic viscosity \citep{ShakuraSunyaev1973} with $\nu=\alpha H^2\Omega$. Different $\alpha$ are adopted to explore the effect of accretion on the migration torque for the planet. For PPDs, the disk viscosity is usually on the order of $\alpha\sim10^{-3}$, but it could be much higher for the innermost region close to the star where the disk temperature is high enough to ionize the gas, and could be more plausible for disks around AGN where stellar-mass objects can be embedded in disks. In this sense, we first focus on the high viscosity case with $\alpha=0.04$ in our 3D simulation and explore different $\alpha$ ranging from $4\times10^{-2}$ to $3\times10^{-4}$ in our 2D simulations.
Such a high viscosity in our 3D local isothermal simulations cannot develop any vertical shear instability as it will be damped completely \citep{Nelson2013}, removing this possible complexity from our simulations. 
A steady accretion rate across the disk is established with the temperature and surface density gradient imposed initially mentioned above.

We also perform several 2D hydrodynamical simulations for parameter survey over different viscosities. In these cases, we choose a 2D cylindrical coordinate 
system $(r,\phi)$.
The disk has an initial profile of 
$\Sigma(r)=\Sigma_{0}\left(\frac{r}{r_{0}}\right)^{-0.5}$,
where $\Sigma_{0}$ is the initial surface density at $r_{0}$ with a natural unit of $M_*/r_{0}^2$. 
This surface density profiles are chosen to match the 3D simulation parameters with $\rho_{0}=\Sigma_{0}/\sqrt{2\pi} H_{0}$, where $H_{0}$ is the disk scale height at $r_{0}$.
The 2D velocity vector of the gas in the inertial frame is $\vec{v}=\left(v_{{r}}, {v}_{\phi}\right)$, and angular velocity is $\Omega=v_{\phi}/r$, although our simulations are carried out in the frame corotating with the planet.

In all cases, we take the natural units of $G=M_{*}=r_{0}=1$. The planet is fixed at a semi-major axis of $a_{\rm p}=r_{0}=1$ with $\Omega_{\rm p} = 
\Omega_0 = 1$.

\subsection{Planetary Accretion}\label{sec:mpdot}

To model the active accretion of the embedded planet, 
we follow previous works and implement a sink hole around the planet (\citealt{Li2023}, see also \citealt{Kley2001,DAngelo2003,Li2021}). 
Accretion is determined by the sink hole radius $r_{\rm a}$, and the removal rate $f$ in unit of local Keplerian frequency $\Omega_{0}$. 
We remove a uniform fraction of mass in every cell around the accretor within $\delta r < r_{\rm a}$ each numerical timestep, 
such that when the density profile within the sink hole settles to a steady state, 
the removal rate converges with the integrated mass flux into the sink hole. 
The accretion rate onto the planet $\dot{m}_{\rm p}$ is computed with

\begin{equation}
\begin{aligned}
\dot{m}_{\rm p} &  =    \int_{\delta r<r_{\rm a}} \eta \Sigma d S,
\end{aligned}
\label{eq:mpdot}
\end{equation}
where $d S$ is the area of the sink cell for 2D simulations
and $\eta$ is the specified gas removal rate. The accretion rates in 3D runs are calculated accordingly with $\dot{m}_{\rm p}  =   \int_{\delta r<r_{\rm a}} \eta\rho dV$, where $dV$ is the volume of the sink cell.
In our fiducial setup, we set $r_{\rm a}=0.1\ R_{\rm H}$ and $\eta=50\Omega_{0}$. 
In an asymptotic limit, $\Sigma$ in sink sphere is $ \propto 
\eta^{-1}$ such that $\dot {m}_{\rm p}$ is independent to the chosen magnitude 
of $\eta$.  In order to verify the dependence on the accretion prescription, we also simulate models with 
$\eta=5\Omega_{0}$, $r_{\rm a}=\epsilon=0.05\ R_{\rm H}$. We find that a smaller accretion/softening radius does not affect the our results significantly, as shown in Table~\ref{tab:para}.
The accretion rate can also be calculated as the mass flux across the sink sphere, 
which is $\dot{m}_{\rm p}=-\oint \Sigma \delta {v}_{r} \delta r d\phi$ for 2D runs, where $\delta\vec{v}_{r}$ is the gas velocity relative to the accretor, and the integration is done in the frame relative to the planet. 
The 3D mass flux onto the accretor in 3D runs can be calculated following to \citet{Li2023} accordingly. 
These two accretion rates are consistent with each other as expected.

We simulate the disk for $\sim10^4$ planetary orbits to ensure the disk reach the viscous steady state for $\alpha > 10^{-4}$.
In all of our simulations, we have fixed the planet's orbital radius and planet mass in order to achieve steady-state measurements.

\subsection{Mesh Refinement and Boundary Conditions}\label{sec:bd}

A static mesh refinement is adopted to resolve the region around the planet. 
For 2D runs, we adopt a base resolution with 128 radial grids spaced  logarithmically uniformly between $r_{\rm min}=0.4\ r_{0}$, $r_{\rm max}=3.0\ r_{0}$, and 512 uniform grids in azimuth. 
We have also tested using a resolution of 256 grids in azimuth, and find no significant change in simulation results. 
For 3D runs, we use a 128 radial grids spaced  uniformly between $r_{\rm min}=0.5\ r_{0}$, $r_{\rm max}=2.5\ r_{0}$, and 512 uniform grids in azimuth for the base level. 
The radial computational domain of the disk is sufficient far away from the planet location of $r_{0}$ to avoid the boundary effect. 
Only half disc above the mid-plane with a vertical domain of $4$ disk scale height is simulated with 16 root grids to save the computation expense after considering the symmetry. 
Three levels of mesh refinement are adopted within the region $\delta r<R_{\rm H}$.
This treatment significantly reduces computation cost and make the simulations extend to $\sim10^{4}$ orbits possible.
For our fiducial case of planet mass $q=0.001$, the Hill sphere can be resolved by about $45$ cells in each dimension.

%The inner edge of the disk is set as $r_{\rm in}=0.4\ r_{0}$.
In the following we mainly focus on the boundary condition for 2D simulations. The boundary conditions for 
the 3D cases are implemented in a very similar way, except for a reflecting boundary in the $\theta$ direction.
We impose a steady mass influx $\dot{m}_{\rm d}$ with 
appropriately extrapolated values of $\Sigma(r)$ and $h(r)$ \citep{Dempsey2020}.
To this end, 
we set the surface density in the ghost zone of the outer boundary based on 
the viscous steady state solution of accretion disk, which is related to $3\pi\Sigma
\nu\ell=(3\pi\Sigma\nu\ell)|_{r_{\rm max}}+(\ell-\ell_{r_{\rm max}})\dot{m}_{\rm d}$, 
where $\ell\equiv (G M_\star r)^{1/2}$ is the specific angular momentum of the disk. This kind of 
outer boundary condition shares some similarities to the interaction simulations in \citet{Miranda2017}.
The radial velocity is set according to $v_{r}=-\dot{m}_{\rm d}/2\pi\Sigma r$.
Such kind of outer boundary is crucial to maintain the correct steady state for the  disk profile.
For the inner boundary, 
we set the $\Sigma$ in the ghost zone to ensure that $\nu\Sigma$ 
is constant and equal to the value at the innermost cell of the computational domain.  
The radial velocity is set at the viscous radial velocity $v_{r}=-3\nu/2r$, 
and the 
azimuthal velocity is set to the pressure-correct Keplerian value. We restrict any mass flow from entering the simulation domain from the inner boundary, which is necessary to produce reliable results of planet accretion as stated in \citet{Li2023}.

The wave killing zones at each radial edge of the disk \citep{deValborroetal2006} is 
applied to damp wave reflections at the radial boundary. Only the radial velocity is 
damped to its azimuthal-averaged value.

\begin{table*}[t]
  \begin{center}
  \caption{\bf Model parameters and simulations results}\label{tab:para}
  \begin{tabular}{l|cccccccccc}
     \hline\hline
     Models & $\alpha$ & $T$ & $r_{\rm a}$ & $\eta$ & $\dot{m}_{\rm d}$ & $\dot{m}_{\rm p}$ & $\dot{a}_{\rm p,grav}/a_{\rm p}$ & $\dot{a}_{\rm p,acc}/a_{\rm p}$& $\dot{a}_{\rm p}/a_{\rm p}$ & Remarks  \\
     & $-$ & ($2\pi/\Omega_{0}$) & ($R_{\rm H}$) & $-$ & 
     ($\Sigma_{0}r_{0}^{2}\Omega_{0}$) &
     ($\Sigma_{0}r_{0}^{2}\Omega_{0}$) & ($\Sigma_{0}r_{0}^{2}\Omega_{0}/M_{*}$) & ($\Sigma_{0}r_{0}^{2}\Omega_{0}/M_{*}$) & ($\Sigma_{0}r_{0}^{2}\Omega_{0}/M_{*}$) &  $-$ \\
     \hline
     A1F & $4\times10^{-2}$ & 3000 & $0.1$ & $50$ & $9.4\times10^{-4}$ & $8.7\times10^{-4}$ & $5.1\times10^{-2}$ & $-5.3\times10^{-3}$ & $4.5\times10^{-2}$ & 3D \\
     A1R5F & $4\times10^{-2}$ & 3000 & $0.1$ & $5$ & $9.4\times10^{-4}$ & $8.6\times10^{-4}$ & $4.7\times10^{-2}$ & $-3.0\times10^{-3}$ & $4.4\times10^{-2}$ & 3D \\
     A1Fna & $4\times10^{-2}$ & 2000 & $-$ & $-$ & $9.4\times10^{-4}$ & $-$ & $-0.23$ & $-$ & $-0.23$ & 3D, no acc \\
     
     A1 & $4\times10^{-2}$ & 10000 & $0.1$ & $50$ & $9.4\times10^{-4}$ & $7.9\times10^{-4}$ & $3.2\times10^{-2}$ & $2.3\times10^{-2}$ & $5.5\times10^{-2}$ & 2D  \\
     A1R5 & $4\times10^{-2}$ & 10000 & $0.1$ & $5$ & $9.4\times10^{-4}$ & $7.8\times10^{-4}$ & $4.4\times10^{-2}$ & $-2.5\times10^{-3}$ & $4.2\times10^{-2}$  & 2D  \\
     A1R5s & $4\times10^{-2}$ & 10000 & $0.05$ & $5$ & $9.4\times10^{-4}$ & $7.4\times10^{-4}$ & $4.9\times10^{-2}$ & $1.4\times10^{-4}$ & $4.9\times10^{-2}$  & 2D  \\
     A2 & $1\times10^{-2}$ & 10000 & $0.1$ & $50$ & $2.4\times10^{-4}$ & $2.0\times10^{-4}$ & $8.5\times10^{-2}$ & $-6.6\times10^{-4}$ & $7.9\times10^{-3}$ & 2D  \\
     A3 & $3\times10^{-3}$ & 40000 & $0.1$ & $50$ & $7.1\times10^{-5}$ & $6.1\times10^{-5}$ & $4.8\times10^{-4}$ & $-1.0\times10^{-3}$ & $-6.2\times10^{-4}$ & 2D  \\
     A4 & $1\times10^{-3}$ & 40000 & $0.1$ & $50$ & $2.4\times10^{-5}$ &$1.8\times10^{-5}$ & $-2.6\times10^{-3}$ & $-3.0\times10^{-4}$ & $-2.9\times10^{-3}$ & 2D  \\
     A5 & $3\times10^{-4}$ & 60000 & $0.1$ & $50$ & $7.1\times10^{-6}$ & $5.1\times10^{-6}$  & $-2.1\times10^{-3}$ & $-7.1\times10^{-5}$ & $-2.2\times10^{-3}$ & 2D  \\

    \hline\hline
   \end{tabular}
   \end{center}
   
   \tablecomments{The planet mass is fixed as $q=0.001$, and the disk aspect ratio is $h_{0}=0.05$ for all models, while the disk viscosity parameter $\alpha$ may vary for different models. 
   The softening radius $\epsilon$ is kept the same as the accretion radius $r_{\rm a}$ for all accreting models except for model \texttt{A1Fna} where $\epsilon=0.1\ R_{\rm H}$. 
   $T$ is the simulation time in unit of the orbital timescale at $r_{0}$. 
   $\dot{m}_{\rm d}$ is the disk accretion rate supplied from the outer boundary. 
   A positive (negative) $\dot{a}_{\rm p}/a_{\rm p}$ means that the planet migrate outward (inward) with time. 
   }
  
\end{table*}

\subsection{Diagnostics}\label{sec:diag}

%We calculate the gravitational torque and accretion torque on the embedded planet in the following. 
The gravitational interaction between the disk material and the planet drive the planet to migrate. 
Another component that modifies the planet orbital dynamics comes from the accreted material onto the planet, 
although we do not add this mass and angular momentum onto the accreting planet actively \citep[e.g.,][]{Li2021b,LiR2024}.
The gravitational force on the planet in 2D simulations is 

\begin{equation}
\vec{F}_{\rm grav}   =    \int\Sigma \nabla \Phi_{\rm p} {\rm d}S,  \\
 %z5&=  \Sigma \frac{ \partial \Phi_{\rm p}}{\partial \phi},
\label{eq:fgrav}
\end{equation}
where $\Phi_{\rm p}$ is the gravitational potential of the planet. The gravitational force for the 3D simulations can be calculated by replacing $\Sigma$ and $dS$ with $\rho$ and ${\rm d}V$, respectively.
The torque due to the gravitational forces are $\vec{\Gamma}_{\rm grav}=\vec{r}_{\rm p}\times\vec{F}_{\rm grav}$, and $\dot{\ell}_{\rm p,grav}=\Gamma_{\rm grav}/m_{\rm p}$.
The force associated with accretion is

\begin{equation}
\vec{F}_{\rm acc}   =   \int_{\delta r<r_{\rm a}} \vec{v}\ {\rm d}\dot{m}_{\rm p},  \\
\label{eq:facc}
\end{equation}
where $\vec{v}$ is the fluid velocity in the inertial frame. 
The accretion torque is $\vec{\Gamma}_{\rm acc}= m_{\rm p}\dot{\vec{\ell}}_{\rm p}+\dot{m}_{\rm p}\vec{\ell}_{\rm p}\equiv \vec{r}_{\rm p}\times\vec{F}_{\rm acc}$, $\vec{\ell}_{\rm p}\equiv \vec{r}_{\rm p}\times \vec{v}_{\rm p}$ is the specific angular momentum of the planet. We thus have
$m_{\rm p}\dot{\vec{\ell}}_{\rm p,acc} \simeq \vec{r}_{\rm p}\times\int_{\delta r<r_{\rm a}} (\vec{v}-\vec{v}_{\rm p})\ {\rm d}\dot{m}_{\rm p}$.

In most cases, the evolution of $a_{\rm p}$ is dominated by the gravitational torque. To diagnostic the origin of this component, we can calculate the cumulative gravitational torque $\Gamma_{{\rm grav},<r}$, which reads as

\begin{equation}
\begin{aligned}
 \Gamma_{{\rm grav},<r} &
=  \int_{r_{\rm in}}^{r}\int_{\rm \phi } \Sigma \frac{ \partial \Phi_{\rm p}}{\partial \phi} r dr d \phi,
\end{aligned}
\label{eq:Gamma_gravcum}
\end{equation}
The total gravitational torque on the planet is thus $\Gamma_{\rm grav}=\Gamma_{{\rm grav},<r_{\rm max}}$.

The semi-major axis evolution of the planet is actually linked to the specific angular momentum change rate of the planet when it is on a \textit{circular} orbit, i.e., $e_{\rm p}=0$, which is

\begin{equation}
\frac{\dot{a}_{\rm p}}{a_{\rm p}}= 2\frac{\dot{\ell}_{\rm p}}{\ell_{\rm p}}-\frac{\dot{M}_{*}}{M_{*}}.
\label{eq:apdot2}
\end{equation}
where $\ell_{\rm p}=\sqrt{GM_*a_{\rm p}}$ is the specific angular momentum for the circular planet, $\dot{\vec{\ell}}_{\rm p}=\vec{r}_{\rm p}\times\dot{\vec{r}}_{\rm p}$ is determined by the external torque from the gravitational and accretional force. 
Specifically, $\dot{\ell}_{\rm p}=\dot{\ell}_{\rm p,grav}+\dot{\ell}_{\rm p,acc}$.
The second term 
is usually much smaller than the first term as confirmed by our simulations.

\begin{figure*}
\centering
\includegraphics[width=0.33\textwidth,clip=true]{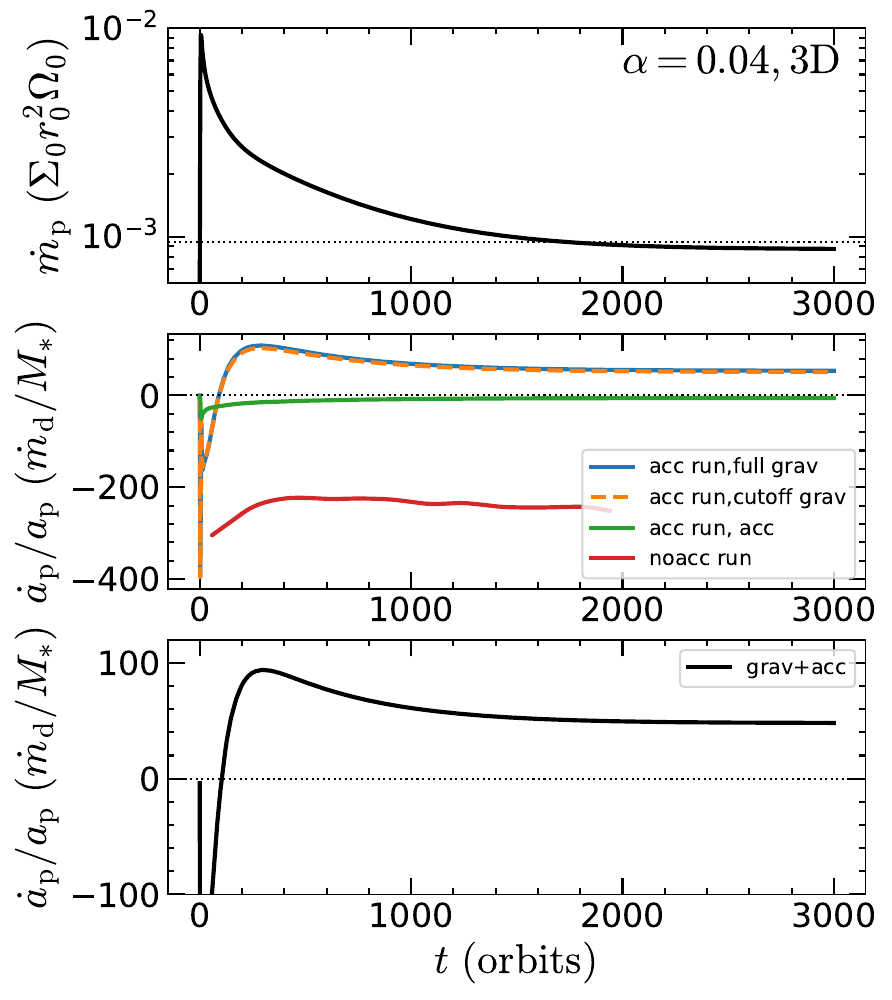}
\includegraphics[width=0.31\textwidth,clip=true]{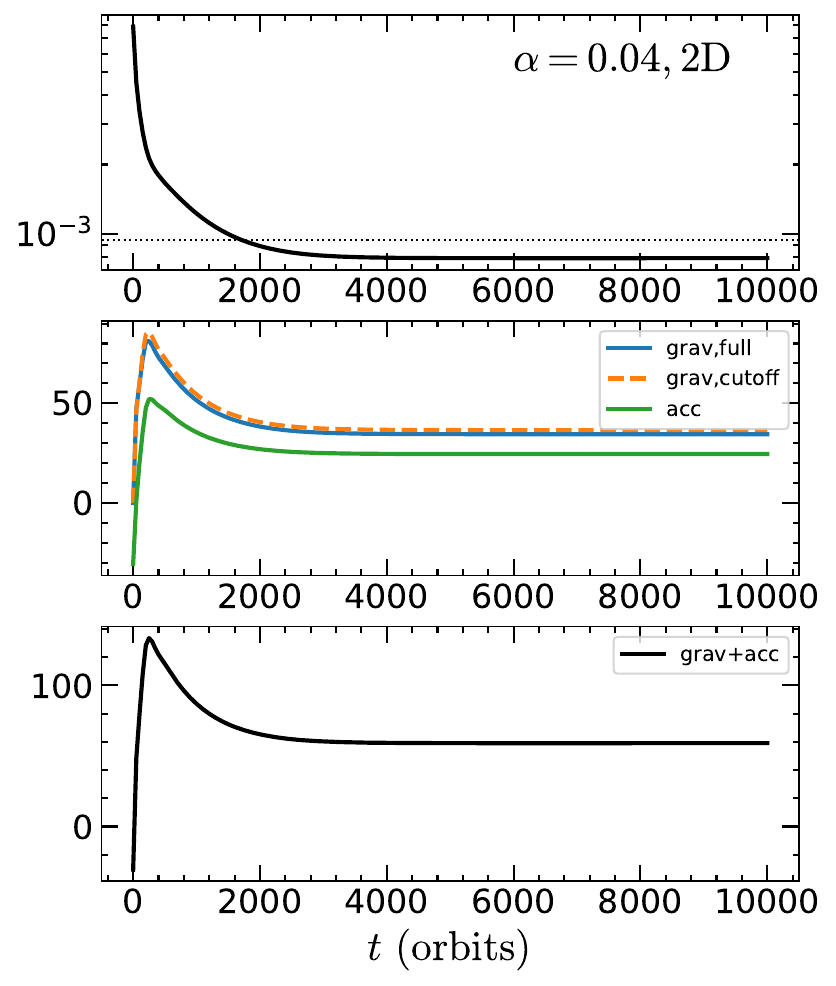}
\includegraphics[width=0.32\textwidth,clip=true]{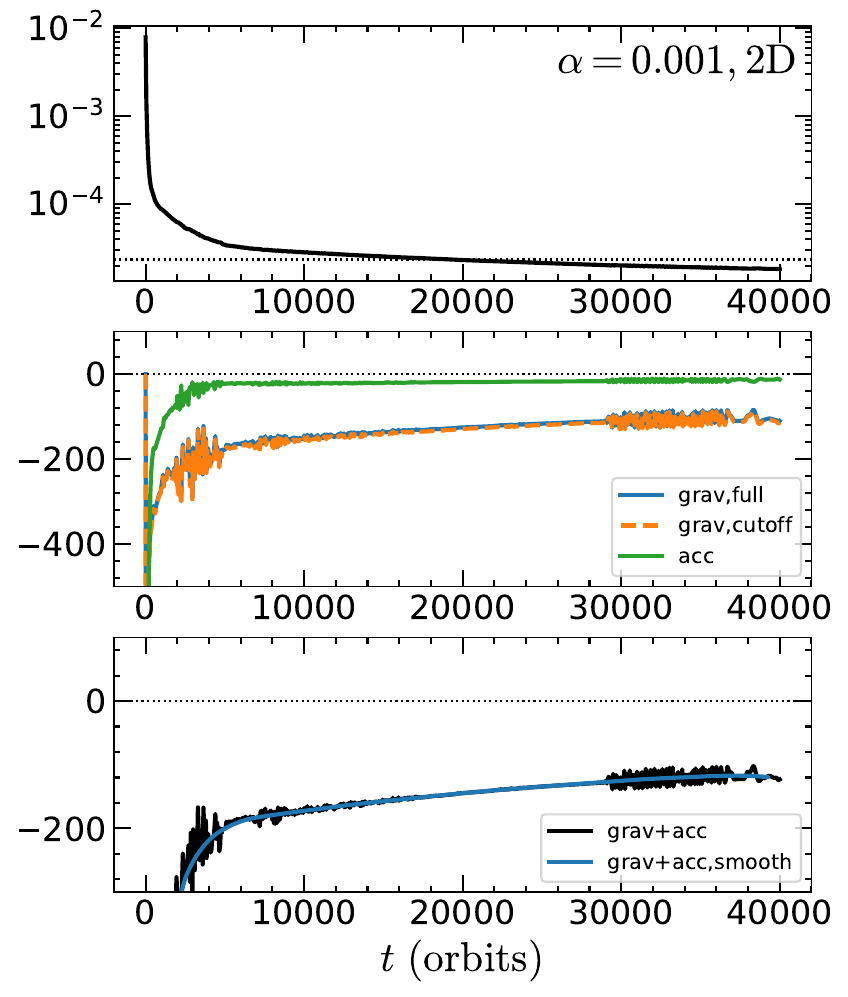}
\caption{(Left Column): The evolution of planetary accretion rate (upper left), semi-major axis evolution $\dot{a}_{\rm p}/a_{\rm p}$ due to the gravitational and accretion torque (middle left) and their sum (lower left) for $\alpha=0.04$ in 3D simulations (model \texttt{A1F}). The dotted line in the upper panel shows the disk accretion supplied steadily from the outer boundary. The blue and green lines in the middle panel show $\dot{a}_{\rm p}/a_{\rm p}$ contributed by the gravitational and accretion torques,  and dashed lines excluded the torque within the sinkhole radius around the planet, which is almost the same the blue solid line. 
The time-averaged $\dot{a}_{\rm p}/a_{\rm p}$ due to the gravitational torque for the non-accreting planet is shown as the red line in the middle panel, which is negative.
(Middle Column):  Same as the left column except for 2D simulations with $\alpha=0.04$ (model \texttt{A1}).  
The total $\dot{a}_{\rm p}/a_{\rm p}$ is consistent with 3D accreting case. 
(Right Column): Same as the middle column except for $\alpha=0.001$ in 2D (model \texttt{A4}). The blue line in the lower panel shows the time averaged $\dot{a}_{\rm p}/a_{\rm p}$ from the total torque.
}
 \label{fig:adot}
\end{figure*}

\section{Results}\label{sec:results}

A summary of model parameters and simulation results is shown in Table~\ref{tab:para}. We mainly explore the effect of disk viscosity on the accretion and migration dynamics.

\subsection{High Viscosity Case}\label{sec:higvis}

We first explore the case with $h_0 = 0.05$ and a very high viscous 
parameter $\alpha=0.04$ in a 3D simulation. This case is labeled as 
model \texttt{A1F} in Table~\ref{tab:para}. At the disk's outer boundary,  
$r_{\rm max}=2.5\ r_{0}$, the viscous time scale $t_{\rm vis} \simeq 4000$ 
planetary orbits. (Note that throughout the paper, time is measured 
in unit of planet's orbital period at $r_{\rm p}=r_{0}$.)
In order to understand the role of planet accretion on the dynamical evolution 
of the planet, 
we evolve the global disk long enough (3000 planetary orbits)
to ensure the disk reach the quasi-steady state. The time evolution of planet 
accretion rate is shown in the upper panel of the left column of Figure~\ref{fig:adot}. 
The disk accretion rate supplied from the outer boundary ${\dot m}_{\rm d}$ is shown 
as horizontal dotted line. After an initial burst of planetary accretion, 
the planetary accretion rate ${\dot m}_{\rm p}$ gradually decreases towards 
a steady rate of $\sim8.7\times10^{-4}\ \Sigma_{0}r_{0}^{2}\Omega_{0}$, which 
is slightly lower than the disk accretion rate at the outer boundary ${\dot m}_{\rm d}$.
In both our 3D (top-left panel) and 2D (top-middle panel) simulations, it can be clearly 
seen that the planetary accretion rate ${\dot m}_{\rm p}$ barely evolves after 
$2000$ orbits. This is simply because the disk supply rate to 
the vicinity of the planet  has reached a quasi-steady state. 
Due to accretion onto the planet, the inner disk is severely depleted with a 
mass flux much smaller than ${\dot m}_{\rm d}$. More discussions about the quasi-steady state are presented in the Appendix~\ref{app:amf}.

\begin{figure*}
\centering
\includegraphics[width=0.45\textwidth,clip=true]{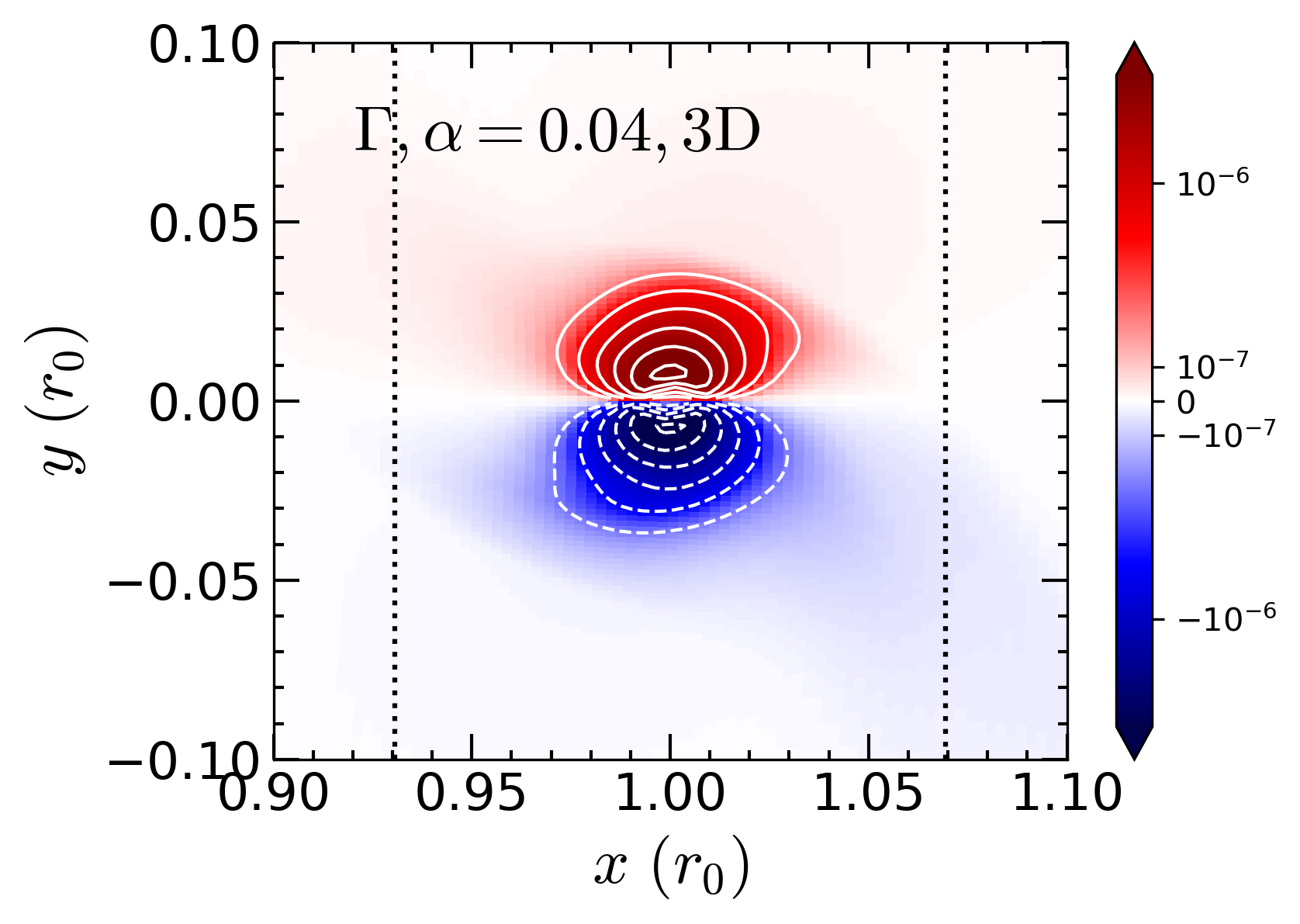}
\includegraphics[width=0.45\textwidth,clip=true]{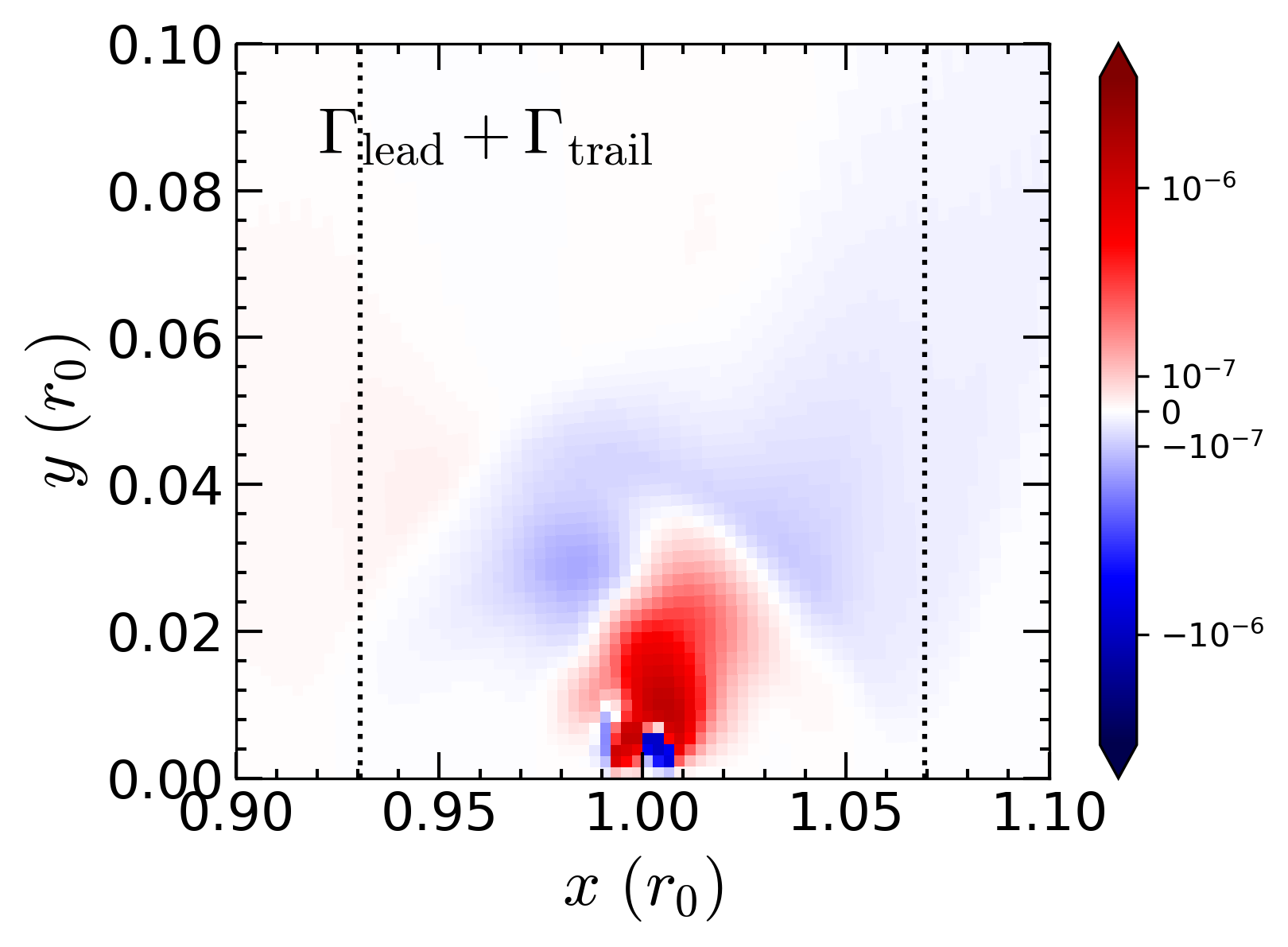}
\includegraphics[width=0.45\textwidth,clip=true]{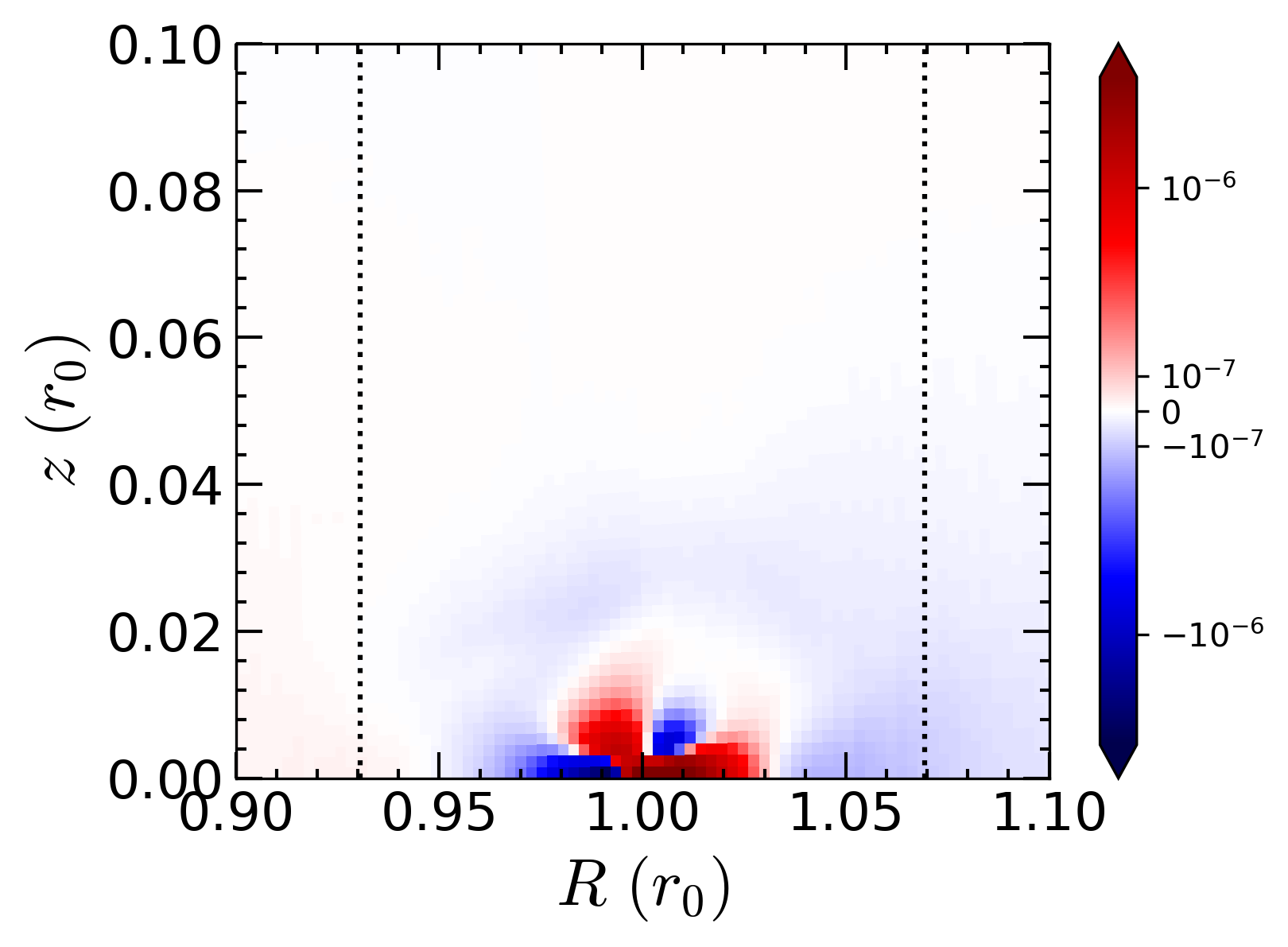}
\includegraphics[width=0.45\textwidth,clip=true]{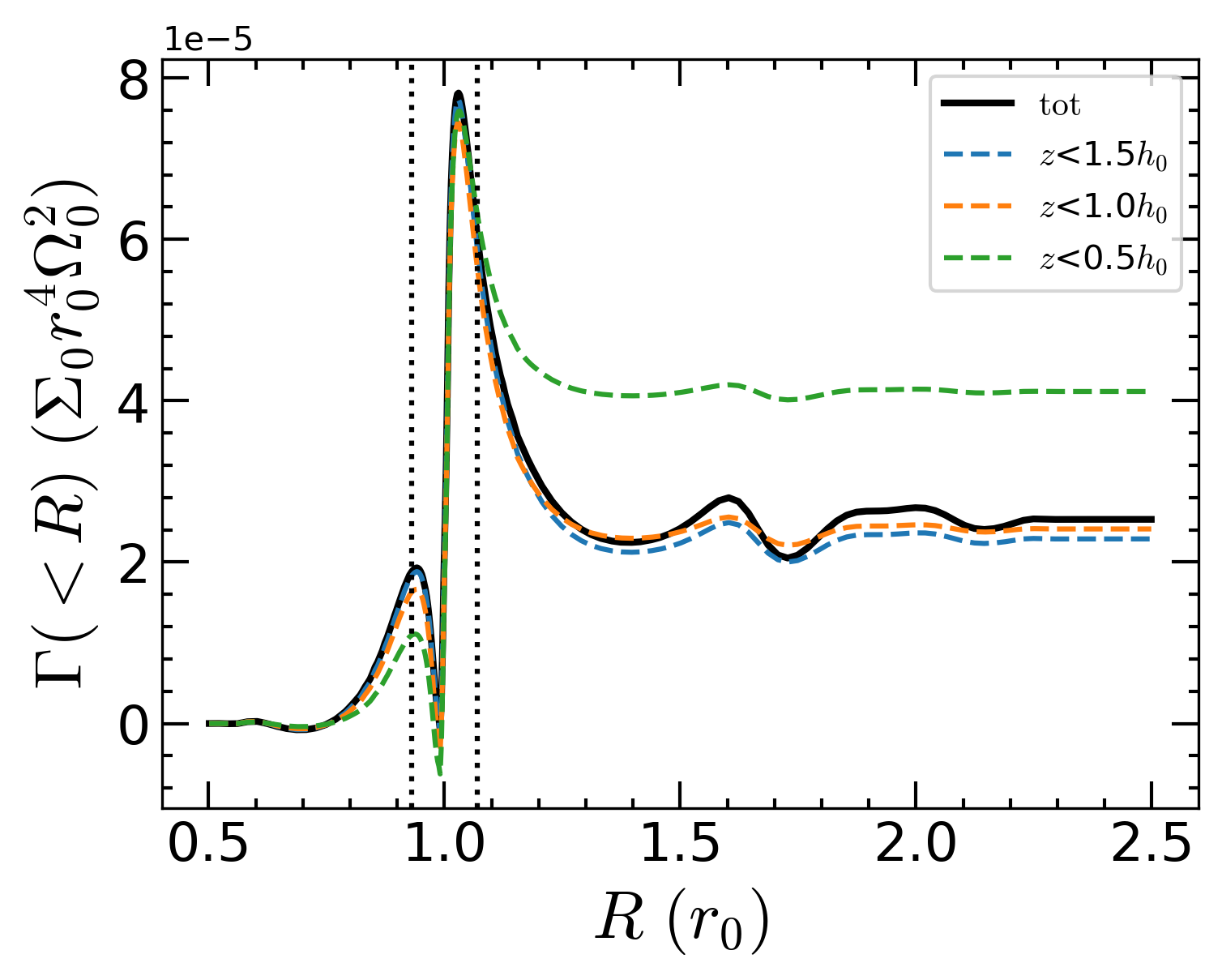}
\caption{Spatial distribution of gravitational torques on the accreting planet for the 3D simulation (model \texttt{A1F}). The upper left panel shows the torque around the planet, with the contour represented by solid and dashed lines being the same magnitude of the torques but with different signs. The upper right panel shows the net torque by summing the leading (upper) horseshoe and trailing (lower) horseshoe region, i.e., $\Gamma(r,\phi)+\Gamma(r,-\phi)$.
The lower left panel shows the azimuthally integrated vertical distribution of the torque. The lower right panel shows the cumulative radial torque distribution integrated within different heights $z$. The total torque can be obtained from the value sufficient large $r \gg r_{\rm p}$. The vertical dotted lines correspond to the region of $1R_{\rm H}$ from the planet.
The planet is located in $(x,y)=(1.0,0.0)r_{0}$ (upper panels) and $(R,z)=(1.0,0.0)$ (lower left panel). It can be seen that the positive torque mostly comes from the co-orbital region.}
 \label{fig:torque_higv_3d}
\end{figure*}

\subsubsection{Accretion and Migration Dynamics}

The time evolution of semi-major axis $\dot{a}_{\rm p}/a_{\rm p}$ due to the 
gravitational torque and accretion torque for the $\alpha=0.04$ 3D model 
\texttt{A1F} is shown in the middle-left column of Figure~\ref{fig:adot}. 
With accretion included, the planet $\dot{a}_{\rm p}/a_{\rm p}$ due to the 
gravitational torque on the planet becomes positive (blue line). 
The 
$\dot{a}_{\rm p}/a_{\rm p}$ associated with the accretion torque, although being 
slightly negative, is much smaller than the gravitational part. 
The total $\dot{a}_{\rm 
p}/a_{\rm p}$ is thus positive, as shown in the lower panel in the same column, which 
suggests that the planet will migrate outward. 
As expected, 
$\dot{a}_{\rm p}/a_{\rm p}$ 
reaches an asymptotic limit, suggesting a steady state of the global disk.
As a sink hole is applied in the vicinity of the planet, some fraction of the 
gravitational torque from the sink cell may be overestimated, we therefore remove all 
the torque within the sink cell, and compare the result (torque with cutoff) with full gravitational torque in the middle row of Figure~\ref{fig:adot}. 
These two prescriptions show very similar profiles, 
suggesting the torque contribution 
from the sink cell is negligible.

\subsubsection{What Causes the Outward Migration?}

\begin{figure}
\centering
\includegraphics[width=0.45\textwidth,clip=true]{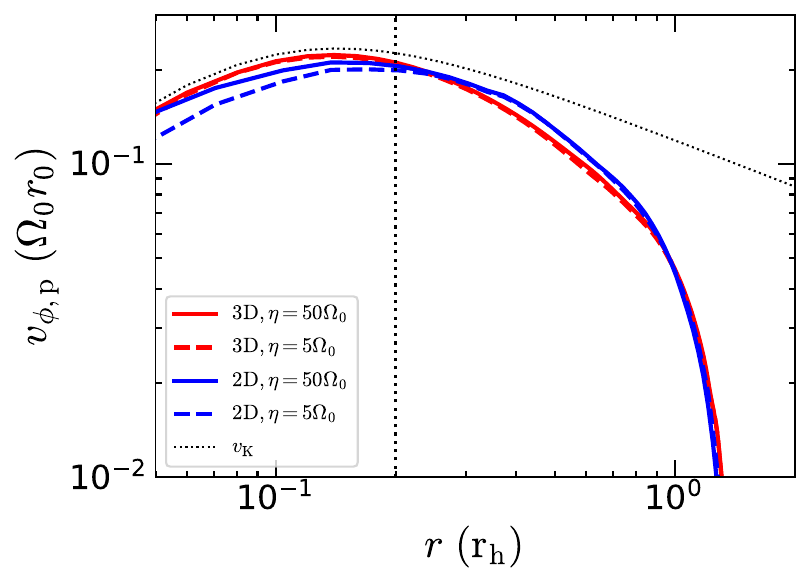}
\caption{Azimuthally averaged rotation velocity around the accretor. Different color lines correspond to different simulations. 
The dotted curve is the Keperian velocity relative to the accretor, 
with the softening effect being included. All the runs are with $\alpha=0.04$. For the 3D runs, the rotation curve is for the mid-plane. 
The vertical dotted line denotes the region within which the gas rotational velocity converges to the local Keplerian value.
We can see that the rotation velocity in the mid-plane almost does not change with the accretion prescription in 3D, however, the velocity profiles vary with the removal rate $\eta$ within the sink cell in 2D runs.}
\label{fig:vp1d_comp}
\end{figure}

\begin{figure*}
\centering
\includegraphics[width=0.45\textwidth,clip=true]{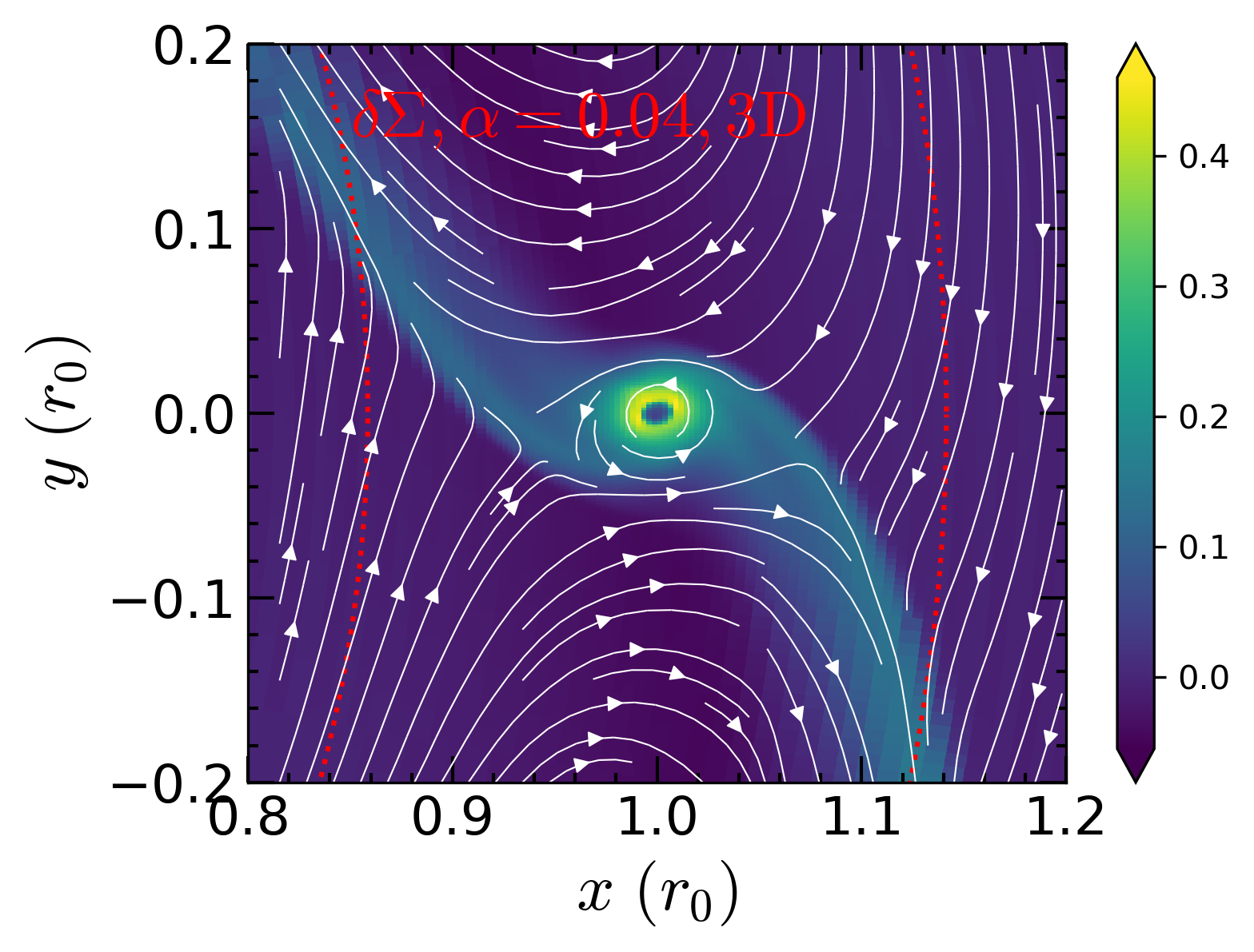}
\includegraphics[width=0.45\textwidth,clip=true]{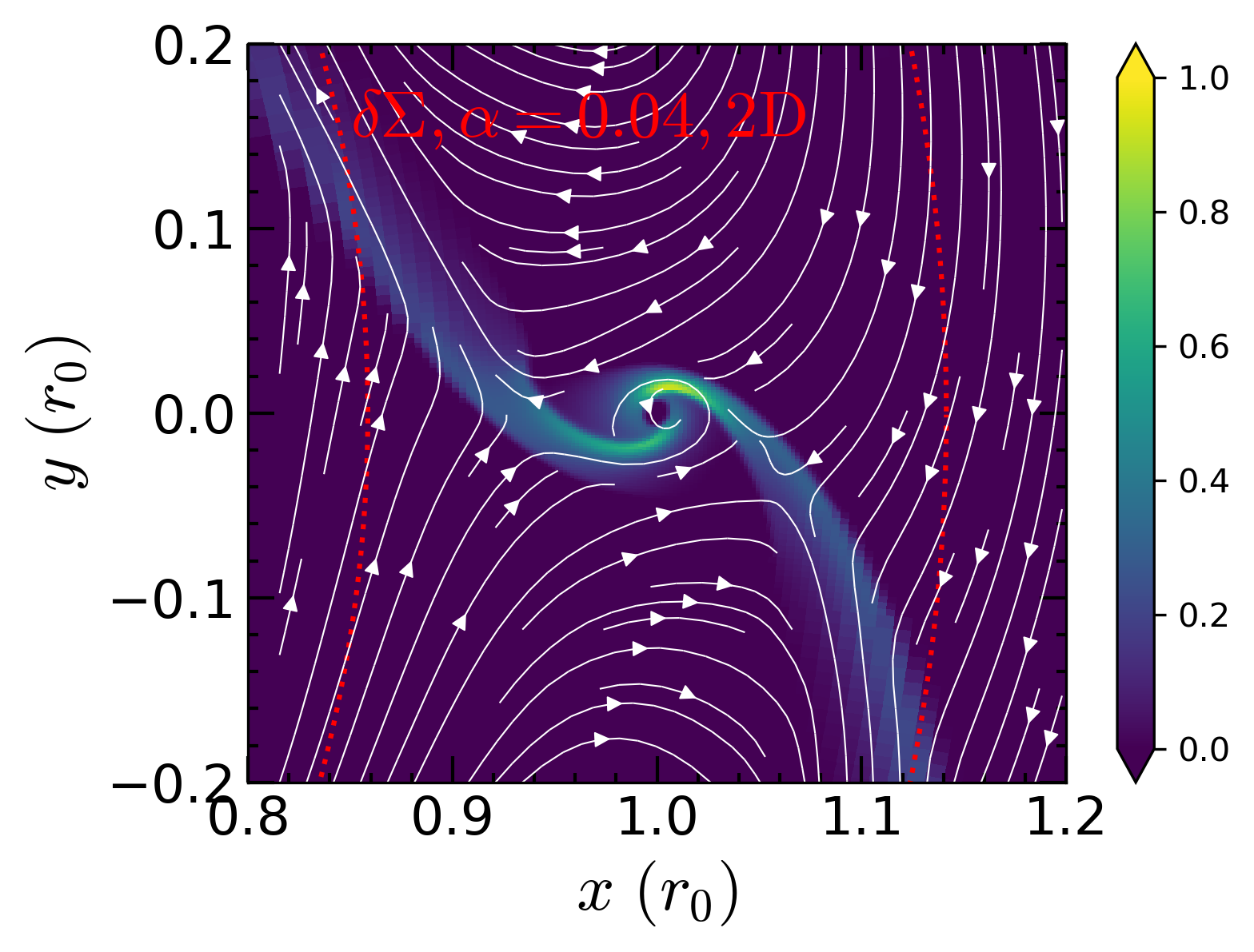}
\includegraphics[width=0.45\textwidth,clip=true]{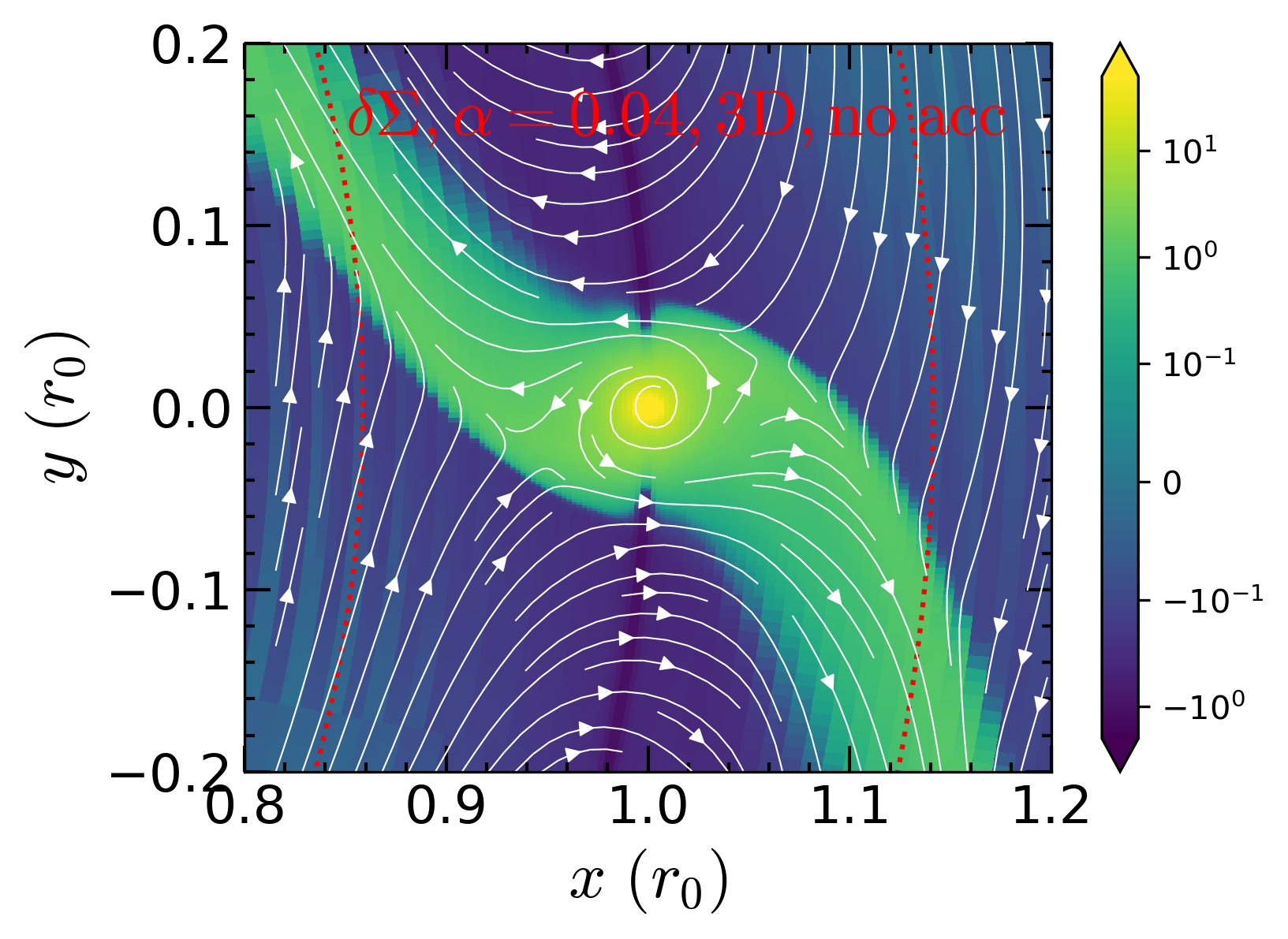}
\includegraphics[width=0.45\textwidth,clip=true]{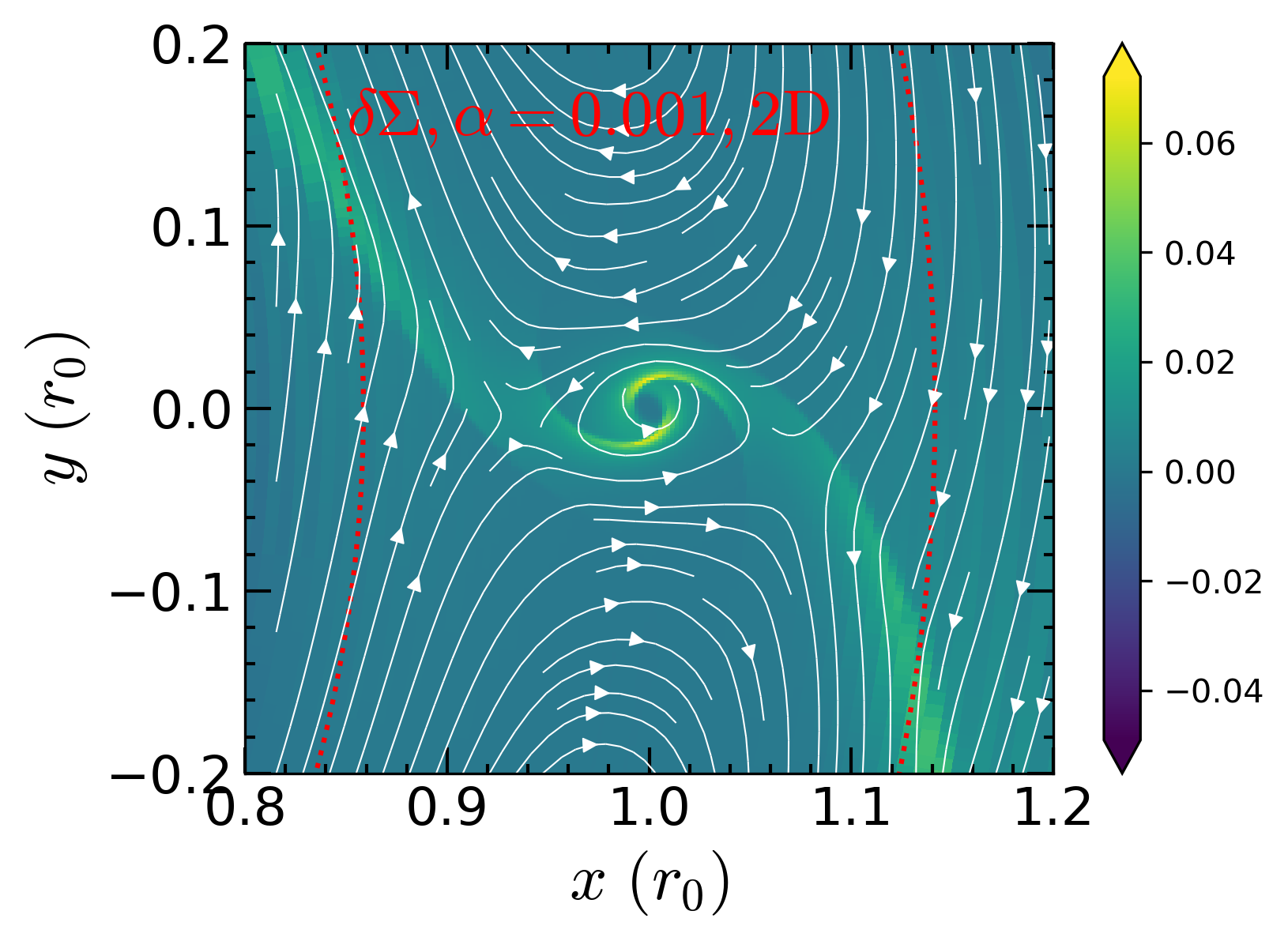}
\caption{Surface density distribution $\delta\Sigma$ around the planet for different runs. The surface densities are subtracted the azimuthal averaged value $\delta\Sigma=\Sigma-\langle\Sigma\rangle$. The overlaid arrows are the streamline line in the co-rotation frame of the planet. The red dotted lines encloses the horseshoe region defined as $x_{\rm s}=a_{\rm p}(q/h)^{1/2}$ \citep[e.g.,][]{KleyNelson2012}. The four runs are 3D accreting planet with $\alpha=0.04$ (upper left, model \texttt{A1F}), 2D accreting planet with $\alpha=0.04$ (upper right, model \texttt{A1}),  3D non-accreting planet with $\alpha=0.04$ (lower left, model \texttt{A1Fnc}), and 2D accreting with $\alpha=0.001$ (lower right, model \texttt{A4}).  The asymmetry between the upper CPD and lower CPD region only shows up in the accreting planet with a high viscosity (upper panels), while the lower two cases present a nearly symmetry density pattern.}
 \label{fig:sigma_streamline}
\end{figure*}

The gravitational torque provides the main contribution for the outward migration. 
Here we decompose the gravitational torque density distribution spatially. 
We do not attempt to distinguish rigorously different contributions from the 
Lindblad region and corotation region, as they usually overlap with each other. 
The vertically integrated gravitational torque distribution around the accreting 
planet $\Gamma(r,\phi)$ around the planet is shown in the upper left panel of Figure~\ref{fig:torque_higv_3d}. 
As expected, the inner and outer spiral arms results in positive and negative torques, respectively, 
and the stronger outer spiral arm results in negative differential Lindblad torque. 

To quantify the torque from different radial band, 
the cumulative radial torque distribution is shown in the lower right panel of Figure~\ref{fig:torque_higv_3d}. 
The negative Linblad torque from the region $r>r_{0}+R_{\rm H}$ is dominates that 
from $r<r_{0}-R_{\rm H}$, as expected from inner disk depletion.
However, the torque integrated from the whole disk 
is positive, and its contribution is dominated by the region very close to the 
planet.

\begin{figure*}
\centering
\includegraphics[width=0.85\textwidth,clip=true]{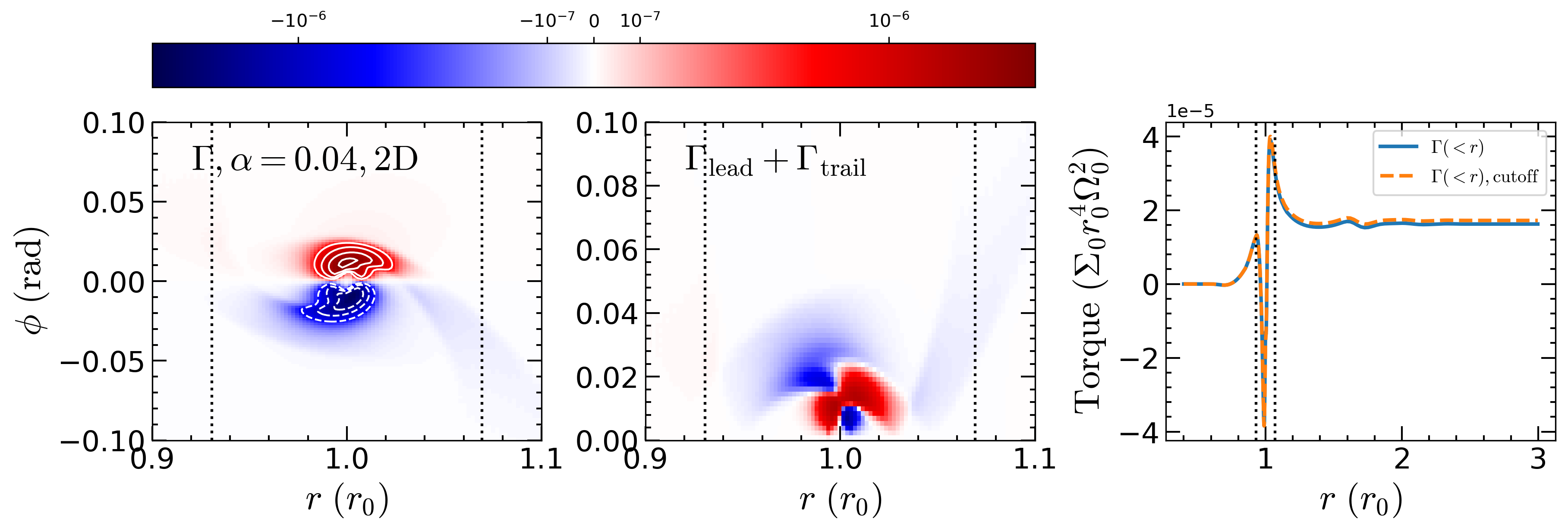}
\includegraphics[width=0.85\textwidth,clip=true]{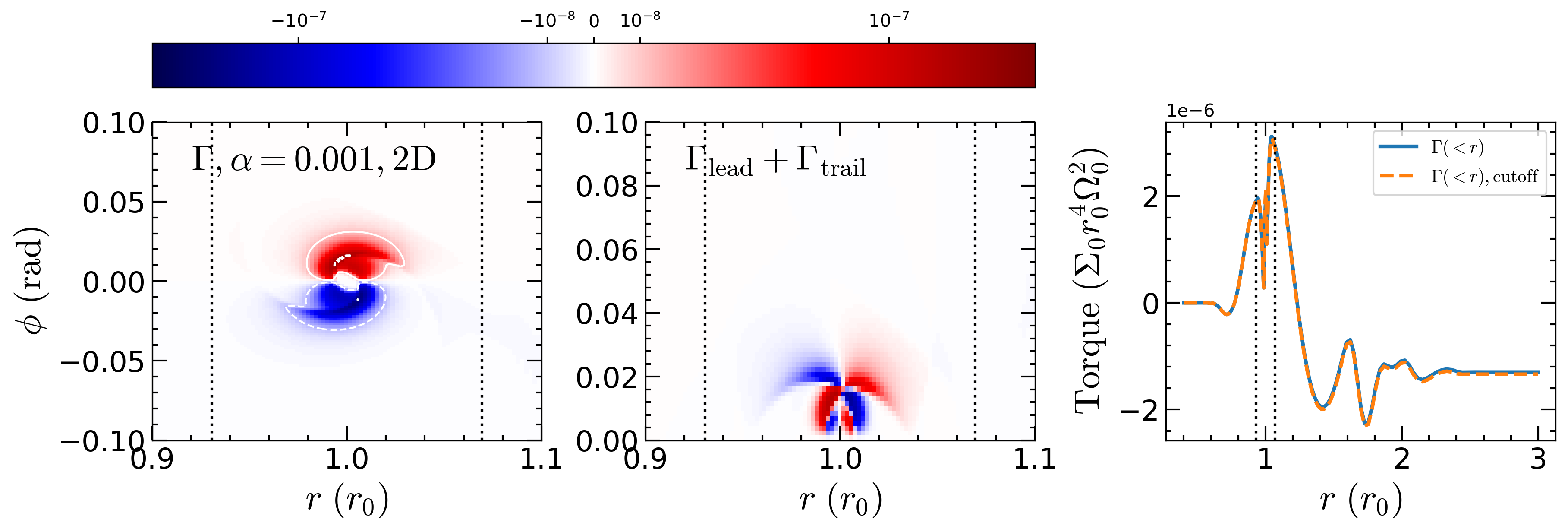}
\caption{Spatial distribution of gravitational torques on the accreting planet for the 2D simulations. 
Upper panel shows the model \texttt{A1} with $\alpha=0.04$, and lower panel is for the model \texttt{A4} with $\alpha=0.001$. 
The left panel shows the torque around the planet, with the contour represented by solid and dashed lines being the same magnitude of the torques but with different signs. The middle panel shows the net torque by summing the leading (upper) horseshoe and trailing (lower) horseshoe region, i.e., $\Gamma(r,\phi)+\Gamma(r,-\phi)$. The right panel show the cumulative radial torque distribution. The total torque can be obtained from the value sufficient large $r \gg r_{\rm p}$. The solid line includes the torque contributed from all the disk material, while the dashed line excludes the contribution from the sink hole region. 
The vertical dotted lines in three panels correspond to the annuli of $1R_{\rm H}$ from the planet.}
 \label{fig:torque_2d}
\end{figure*}

The upper right panel of Figure~\ref{fig:torque_higv_3d} shows the 
vertically integrated, azimuthal differential 
torque around the planet defined as $\Gamma_{\rm asy}=\Gamma(r,\phi)+\Gamma(r,-\phi)$, 
such that only the asymmetric component of the torque is left \citep{Li2021b,Li2022b}. Close to the planet, 
the positive torque from the leading (with $\phi \geq 0$) hemisphere is stronger than the negative torque from the trailing (with $\phi 
\leq 0$) hemisphere. Note that
based on the rotation profile in the frame relative to the planet as shown in Figure~\ref{fig:vp1d_comp}, 
the rotation-supported circumplanetary disk (CPD) only exists within $\delta r\lesssim 0.2-0.3\ R_{\rm H}$ where the azimuthal velocity converges to Keplerian. 
The gas orbiting the planet flows into the CPD and then back out within a finite time, which suggests the gas is not formally bound to the planet \citep[see review by][]{Paardekooper2023}. 
Comparing with Figure~\ref{fig:torque_higv_3d} (also seen in the upper panel in Figure~\ref{fig:torque_2d} discussed below), we observe that the asymmetric contribution of the positive torque comes from the region \textit{outside} of $0.15\ R_{\rm H}$, but not directly from the bound CPD.

The azimuthally integrated torque distribution in the $r-z$ plane is shown in the lower left panel of Figure~\ref{fig:torque_higv_3d}, and the cumulative radial torque distribution from different vertical heights is shown in the lower right panel of Figure~\ref{fig:torque_higv_3d}. Most of the positive gravitational torque is dominated by the region within $z<0.5h_{0}$, 
while there are some negative torque contributions from higher altitude. 
Nevertheless, the total gravitational torque contribution from different heights $z$ is always positive due to the strong positive torque close to the CPD region.

Such kind of azimuthal asymmetry is associated with the proximity of the spiral arms to 
the planet, 
which leads to a strong positive torque from the leading hemisphere of the Hill sphere connecting with the 
CPD region.
This inference is confirmed by the cumulative radial torque 
distribution in the lower right panel of Figure~\ref{fig:torque_higv_3d}.
The asymmetric pattern between the leading and trailing hemisphere close to the CPD region can be identified from the surface density distribution 
around the planet, as shown in the left panel of Figure~\ref{fig:sigma_streamline}. 
This is similar to the hydrodynamical simulations of circumbinary disk and embedded binary in AGN disk where the circum-single disk can contribute most of
gravitational torques for the binary system \cite[e.g.,][]{Tiede2020,Li2021b,Li2022b}. It should be expected that this is different from the bound material \textit{within} the CPD region where the direct gravitational torque acting on the planet itself can be impossible \citep{Papaloizou2007,Crida2009,Terquem2011}. For example, \citet{Crida2009} identified a bound CPD with its size of $\sim 0.5\ R_{\rm H}$ by adopting a large softening length of $\sim 0.4\ R_{\rm H}$ for a non-accreting planet.

For an accreting planet embedded in a disk with high viscosity, 
most of the accreted material first loads
onto the horseshoe tracks from the (azimuthally) leading ($\phi > 0$), outer 
($r > r_{\rm p})$ regions of the disk.  After passing the planet, a considerable fraction of the
gas on the streamline gets accreted by the planet.  
In contrast, only a small fraction of disk gas moves away (in azimuth) from the planet to join the trailing ($\phi < 0$) branch of the 
horseshoe tracks. 
The difference between the
leading and trailing horseshoe flows results in an asymmetric distribution of materials feeding the CPD region around the accreting planet.
This phenomenon is similar to the unsaturated horseshoe drag effect \citep{Masset2001}, albeit in that scenario the imbalance is caused by gas
diffusing into the inner ($r < r_{\rm p}$)
regions of the disk before completing a full libration along the horseshoe track. However, in the context of accreting planets the torque imbalance is much larger than the unsaturated corotation torque being produced.
The net gravitational torque from the co-orbital region is largely not affected by whether we 
have included the contribution from the sink cell or not, confirming the discussion 
above (lower-right panel in Figure~ \ref{fig:torque_higv_3d}).

\begin{figure*}
\centering
\includegraphics[width=0.45\textwidth,clip=true]{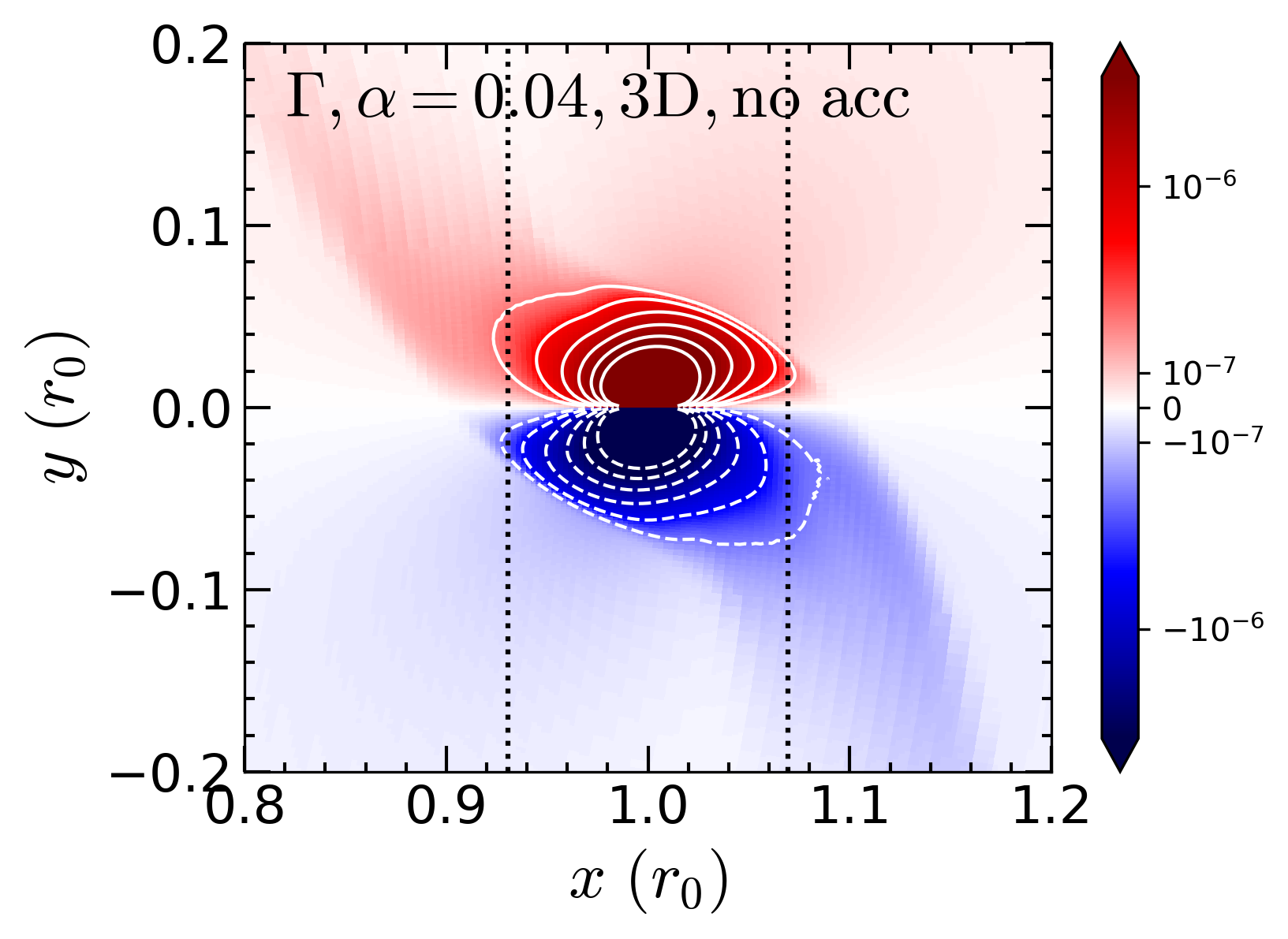}
\includegraphics[width=0.45\textwidth,clip=true]{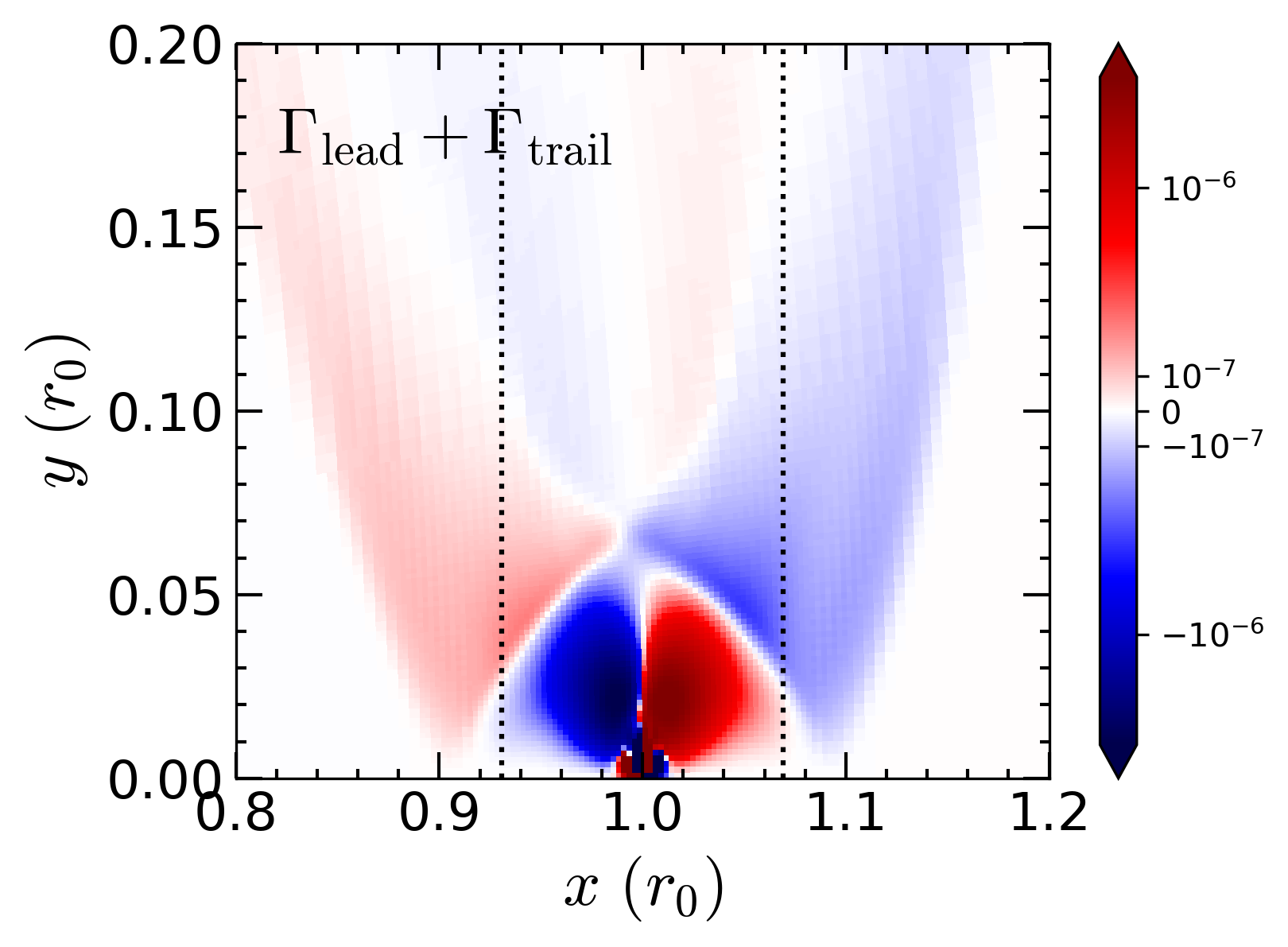}
\includegraphics[width=0.45\textwidth,clip=true]{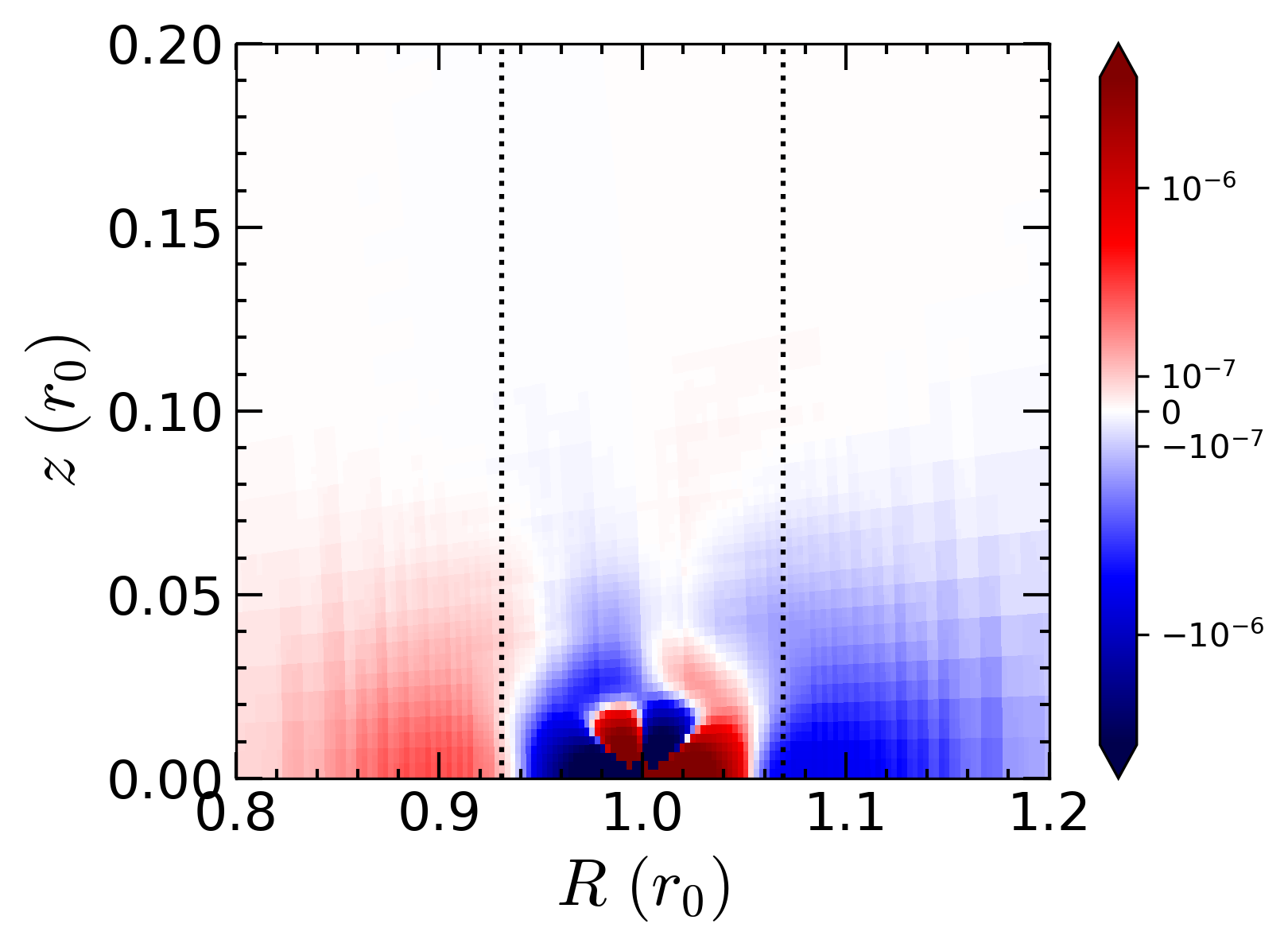}
\includegraphics[width=0.45\textwidth,clip=true]{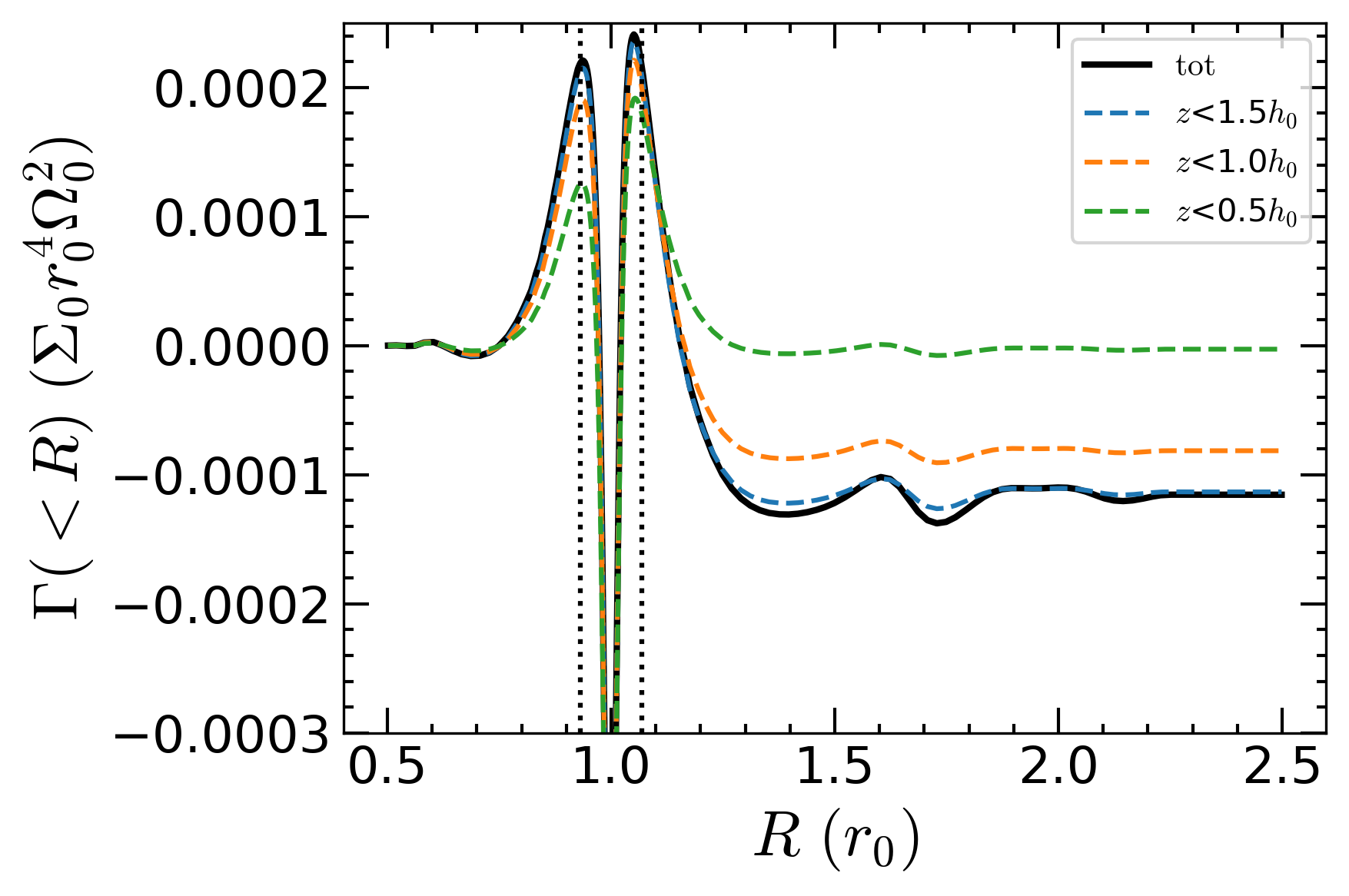}
\caption{Same as Figure~\ref{fig:torque_higv_3d} but for the non-accreting planet. A time-averaged within $100$ orbits is adopted to smooth out the temporal torque oscillation attributed from the co-orbital region.}
 \label{fig:torque_higv_noacc_3d}
\end{figure*}

A similar 3D run but without accretion is carried out to compare with the accreting case. 
The $\dot{a}_{\rm p}/a_{\rm p}$, nearly exclusively contributed by gravitational torque, is also shown in the left column of Figure~\ref{fig:adot}. 
Azimuthal symmetry in density along the streamlines is restored (lower left panel, Figure~\ref{fig:sigma_streamline}) and the negative $\dot{a}_{\rm p}/a_{\rm p}$ recovers the classical inward migration for non-accreting planet. 
The $\dot{a}_{\rm p}/a_{\rm p}$ is also consistent with the classical transitional migration rate, which is $\simeq-2.0q/h_{0}^{2}\Sigma r^{2}\Omega/M_{*}\simeq-200 \dot{m}_{\rm d}/M_{*}$ \citep[e.g.,][]{Tanaka2002,Paardekooper2010,Kanagawa2018}, 
when there exists no deep gap around the planet for our high viscosity adopted here.

Similar to the accreting case, the time-averaged torque distribution for the 3D non-accreting planet is shown in Figure~\ref{fig:torque_higv_noacc_3d}.
The torque magnitude is now much stronger,
simply due to mass pile up around the non-accreting planet. More importantly, the upper and lower co-orbital region show very similar pattern, as shown in the upper right of Figure~\ref{fig:torque_higv_noacc_3d}. As a result, the net torque from the co-orbital region is nearly cancelled out, leading to nearly zero torque. 
The total torque is now dominated by the differential Lindblad torque, which is negative as expected, as shown in the lower right panel of Figure~\ref{fig:torque_higv_noacc_3d}.
Due to the dense material pile up in the vicinity of the planet, 
the torque distribution is more vertically extended as shown in the lower left panel, 
and confirmed quantitatively by the lower right panel in the same figure.

\subsubsection{Comparison with 2D Simulations}

In order to model the planet accretion and CPD dynamics, a small softening/sink cell radius is needed in 3D simulations. 
In addition, the 3D disk does not require a large softening to mimic the effect of the vertical extension of the disk, which is different from previous 2D simulations to study the migration of non-accreting planets.
While a small softening/sink cell radius is justified for the 3D simulation, it is usually quite computationally unfeasible to simulate the disk to a viscous steady state, especially for a low viscosity disk.
To this end, a 2D simulation with the same model parameters and accretion/softening prescription are thus carried out to compare the migration dynamics shown in the 3D case.

The planetary accretion rate and semi-major axis evolution for model \texttt{A1} are shown in middle column of Figure~\ref{fig:adot}. 
We evolve the system to 10000 orbits, 
which is much longer than that we can achieve with the 3D simulation.
The measured accretion rate and $\dot{a}_{\rm p}$ actually does not show significant evolution after 3000 orbits. 
The planetary accretion rate starts to decline below the disk supply rate from the outer boundary at 2000 orbits, similar to 3D case. More importantly, both $\dot{m}_{\rm p}$ and $\dot{a}_{\rm p,grav}/a_{\rm p}$ in 2D simulations are consistent with 3D results, 
although the $\dot{a}_{\rm p,acc}/a_{\rm p}$ associated with accretion is slightly different from that of 3D simulation (see 
further discussions in Section~\ref{sec:acc_eta}). Under the combined gravitational 
and accretion torques, 
the net $\dot{a}_{\rm p}/a_{\rm p}$ is positive and well consistent with the 3D simulation. This agreement suggests the robustness of 
outward migration for the accreting planet found in different simulations.
We can also calculate the $\dot{a}_{\rm p}/a_{\rm p}$ based on the angular momentum flux transfer in the disk, which is shown in the Appendix~\ref{app:amf}. This shows a consistent $\dot{a}_{\rm p}/a_{\rm p}$ measurement, again confirming our results.

Similar to the 3D simulations, the spatial distribution of the gravitational torque is shown in the upper panel of Figure~\ref{fig:torque_2d}. The torque distribution around the accreting planet is plotted in the left panel, 
which shows a similar pattern as the 3D simulation shown in Figure~\ref{fig:torque_higv_3d}. The asymmetric component of the torque is shown in the middle panel of Figure~\ref{fig:torque_2d}. It is obvious that the positive torque from the leading sector of the co-orbital region is stronger than the negative component in the trailing co-orbital counterpart, which results in a net positive torque from the co-orbital region (see also the right panel of Figure~\ref{fig:torque_2d}). 
Such a pattern are qualitatively in agreement with 3D simulations. 
The total gravitational torque from the whole disk also matches the 3D simulations, although the magnitudes of the torque in different regions are different.
Such a discrepancy could come from the vertical extension of the torque in 3D simulations as discussed above.
As in 3D accreting case, the torque removal within the softening radius does not change the total torque acting on the planet.

\begin{figure}
\centering
\includegraphics[width=0.45\textwidth,clip=true]{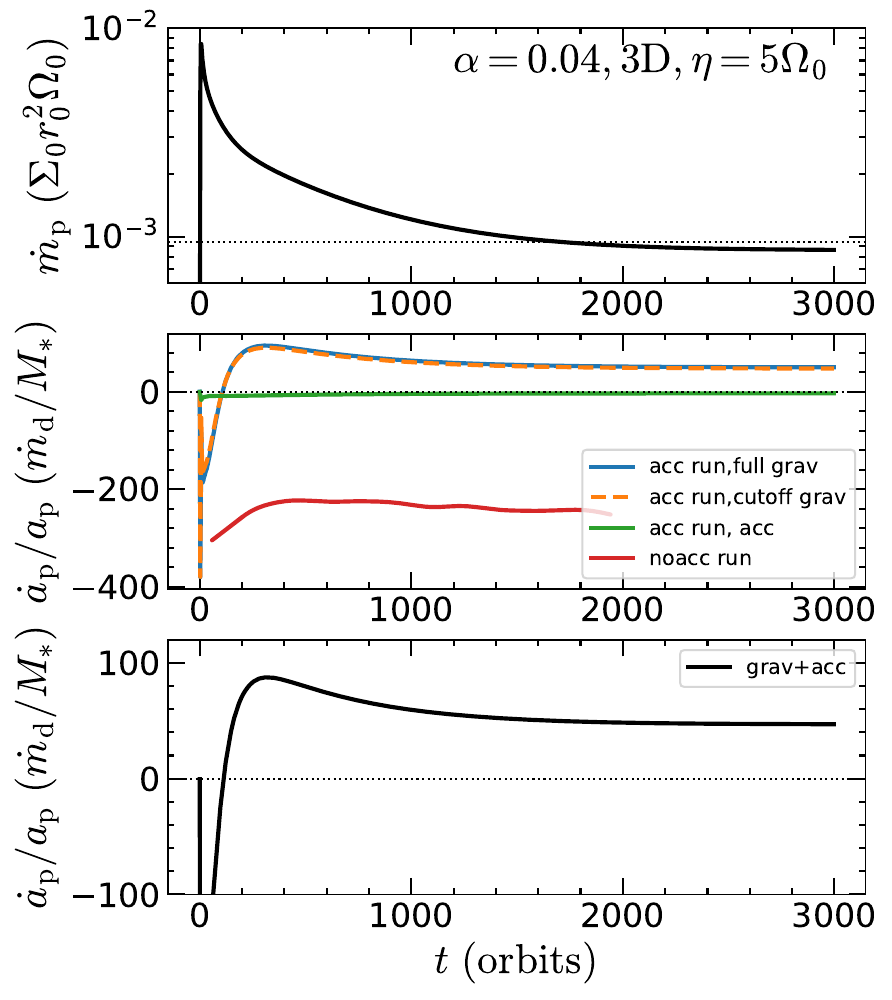}
\caption{Same as the left column of Figure~\ref{fig:adot} except that the removal rate $\eta=5\Omega_{0}$ (model \texttt{A1R5F}).  The $\dot{m}_{\rm p}$ and $\dot{a}_{\rm p}$ are well consistent with the 3D case with $\eta=50\Omega_{0}$, although the accretion torque is also different from the 2D case as shown in Figure~\ref{fig:adot}. Nevertheless, the total $\dot{a}_{\rm p}$ are quite consistent among different simulations.}
 \label{fig:adot_higv_acc2}
\end{figure}

\subsubsection{Dependence on the Sink Radius and Removal Rate }\label{sec:acc_eta}

Thus far, an accretion radius of $r_{\rm a}=0.1\ R_{\rm H}$ and removal rate of $\eta=50$ is adopted to efficiently remove mass within the sink cell, which
produces a sink hole of density around the planet.
%which can ensure the convergence of the planetary accretion rate onto the planet. 
To explore the robustness of removal rates on measurement of migration torques, 
we perform another two simulations with $\eta=5\Omega_{0}$ both in 3D (\texttt{A1R5F}) and 2D (\texttt{A1R5}), 
while keeping other parameters as our high viscosity cases as shown in the first two columns of Figure~\ref{fig:adot}. The model parameters and simulations results are listed in Table~\ref{tab:para}. 
The planetary accretion rate and $\dot{a}_{\rm p}/a_{\rm p}$ profiles for the 3D run are shown in Figure~\ref{fig:adot_higv_acc2}. 
As we can see, 
both the planetary accretion rate and $\dot{a}_{\rm p}/a_{\rm p}$ due to the gravitational torque are well consistent with the case of $\eta=50\Omega_{0}$, 
while the $\dot{a}_{\rm p}/a_{\rm p}$ associated with the accretion part changed by order unity although its effect on total torque is overshadowed by the gravitational counterpart. 
The total $\dot{a}_{\rm p}$ is, nevertheless, positive and quite consistent with that of $\eta=50\Omega_{0}$ in 3D.

However, we note that the accretion torque computed with two values of $\eta$
in the 2D simulations is different from each other, and it is 
also different from that obtained from their 3D counterparts. 
To understand such a dependence in 3D/2D simulations, we plot the 
azimuthally averaged rotation velocity in the CPD 
around the accretor for different runs, as shown in Figure~\ref{fig:vp1d_comp}.
The mass accretion rates onto the planet reach satisfactory convergence for different removal rates. It suggests that the relative velocity ($\vec{v}-\vec{v}_{\rm p}$) in 2D varies with the increasing $\eta$ 
while fixing $\dot{m}_{\rm p}$ since gas surface density decreases with removal rate, which alters the mid-plane pressure support. 
This translates into a variation of the rotational velocity profiles (blue lines) is shown in Figure~\ref{fig:vp1d_comp} 
, which in turn results in the dependence of the accreted angular momentum on $\eta$. 
However, in 3D cases
the accreted angular momentum mostly comes from the mid-plane, 
while the accreted mass flux is dominated by the high latitude (\citealt{Li2023}, also confirmed here). 
The 3D geometry results in a weaker dependence of mid-plane rotation as a function of the mid-plane density gradient or removal rate $\eta$, 
as shown in the mid-plane rotational velocity profiles (red lines) in Figure~\ref{fig:vp1d_comp}.
Nevertheless, 
the total torque between 2D and 3D with different removal rates is still comparable, 
and their minor difference does not change our conclusions. 

To further explore the dependence of migration torques on the sink radius, we carry out another simulation with $r_{\rm a}=0.05\ R_{\rm H}$ and set the softening radius as the same length scale as well. The results labelled as model \texttt{A1R5s} are shown in Table~\ref{tab:para}. It can be seen that compared with model \texttt{A1R5} satisfactory convergence has been made, although the accretion torque still shows some discrepancy.

\subsection{Low Viscosity Case}\label{sec:lowvis}

Up to now, 
all the simulations applied high viscosity parameter of $\alpha=0.04$ corresponding to a short viscous timescale $t_{\rm vis} \propto \alpha^{-1}$. 
Here we explore the dependence of the migration dynamics on the viscosity parameter since the co-orbital region could be sensitive to the disk viscosity. 
We will mainly resort to the global 2D simulations for these low viscosity cases, 
since it is found based on our previous comparisons that the 2D simulations can generally reproduce the accretion rates and migration torques in 3D cases. 
In addition, it is quite computationally unfeasible to carry out a global 3D simulation to reach a steady state with a very low viscosity. 
We prefer to adopt a larger removal rate of $\eta=50$ as to reach the asymptotic convergence state.

Here we present a 2D simulation with $\alpha=10^{-3}$. 
The time evolution profiles for the planetary accretion rate and $\dot{a}_{\rm p}/a_{\rm p}$ are shown in the middle column of Figure~\ref{fig:adot}. Due to the low viscosity, we evolve the system much longer than that of high viscosity case to reach a quasi-steady state, after that $\dot{m}_{\rm p}$ and $\dot{a}_{\rm p}/a_{\rm p}$ does not show any significant evolution. 
The measured $\dot{a}_{\rm p}/a_{\rm p}$ due to the gravitational force becomes negative, regardless whether we have excluded the contribution from the sink cell region, while the $\dot{a}_{\rm p}/a_{\rm p}$ associated with accretion is slightly negative. 
The net $\dot{a}_{\rm p}/a_{\rm p}$ turns out to be negative because of the strong negative $\dot{a}_{\rm p}/a_{\rm p}$ from the gravitational torque. 

Similar to the analysis of the high viscosity case, the spatial torque distribution is shown in the lower panel of Figure~\ref{fig:torque_2d}. 
The azimuthal asymmetry between the leading (upper) co-orbital and trailing (lower) co-orbital region disappears as shown in the lower middle panel of Figure~\ref{fig:torque_2d}. This pattern is also confirmed by the symmetric density distribution as shown in the lower right panel of Figure~\ref{fig:sigma_streamline}. The disappearance of the asymmetry is simply because of the weak viscous diffusion along the horseshoe streamline such that almost all the material from the leading (upper) horseshoe can finally enter the trailing (lower) horseshoe.
As a result, the net torque from the co-orbital region is almost cancelled out, leaving the main torque contribution coming from the differential Lindblad torque, as shown in the lower right panel of Figure~\ref{fig:torque_2d}.

\subsection{Scaling Relation and Transition between Outward and Inward Migration}

\begin{figure*}
\centering
\includegraphics[width=0.8\textwidth,clip=true]{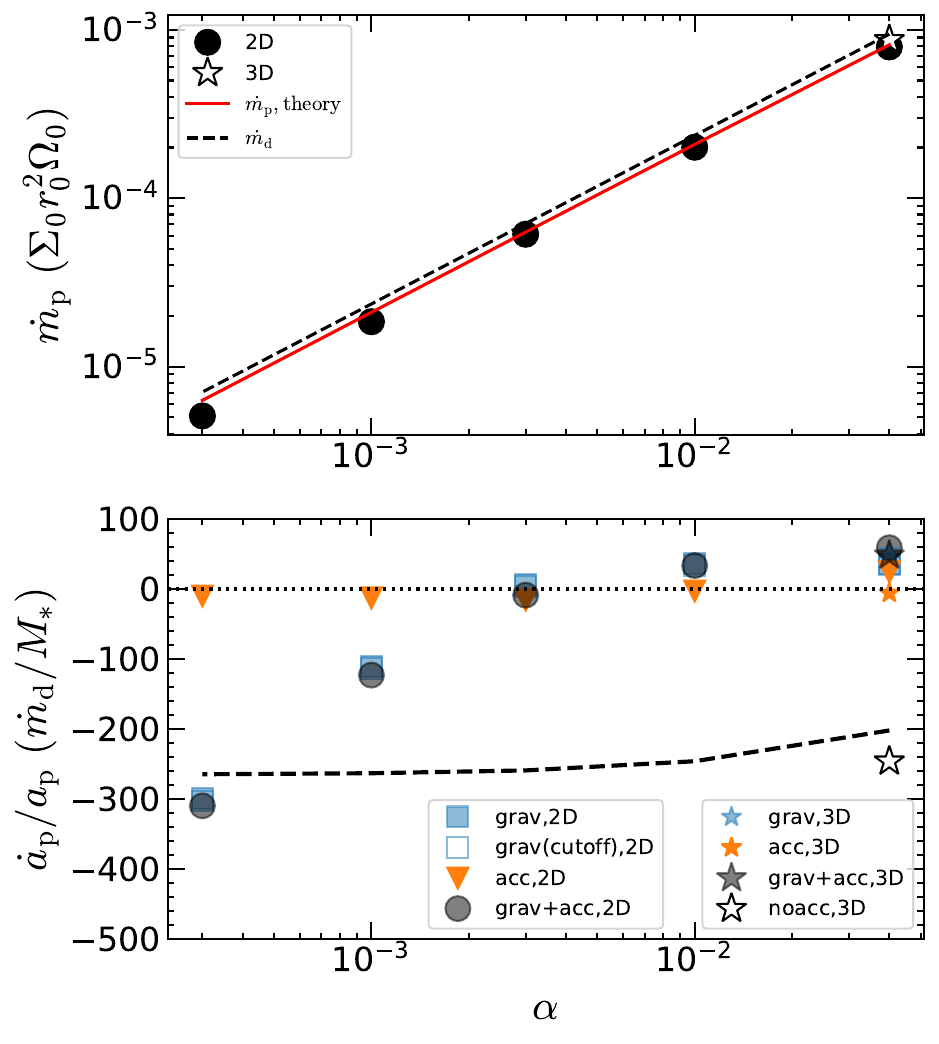}
\caption{Planetary accretion rates (upper panel) and $\dot{a}_{\rm p}/a_{\rm p}$ (lower panel) as a function of viscosity parameter $\alpha$. Upper panel: the red line shows the theoretical accretion rates based on Eq.~\ref{eq:mpdot_th}, which perfectly predicates our simulations results.  Dashed line is the disk accretion rate supplied from the outer boundary. Lower panel: symbols with different colors correspond to different component of $\dot{a}_{\rm p}/a_{\rm p}$, and the symbols of star represent the 3D runs. Only the runs with a removal rate of $\eta=50\Omega_{0}$ for accreting planets in Table~\ref{tab:para} are shown. There is a transition from inward migration to outward migration around $\alpha\sim3\times10^{-3}$. The dashed line shows the classical type I/II migration rates where the disk depletion effect due to gap opening has been considered.}
 \label{fig:adot_alpha}
\end{figure*}

In order to understand the transition from planet outward to inward migration, 
we carry out a few additional runs with different values of viscosity. The results are summarized in Table~\ref{tab:para} and Figure~\ref{fig:adot_alpha}.
In the upper panel of Figure~\ref{fig:adot_alpha}, we show the planetary accretion rate as a function of disk viscosity. 
According to \citet{Li2023}, the theoretical planetary accretion rates are predicted to be 

\begin{equation}
\dot{m}_{\rm p}= \frac{\dot{m}_{\rm H}\dot{m}_{\rm d}}{\dot{m}_{\rm H}+\dot{m}_{\rm d}}.
\label{eq:mpdot_th}
\end{equation}
Here $\dot{m}_{\rm d}$ is the disk accretion rate supplied from the outer boundary, and $\dot{m}_{\rm H}$ is 

\begin{equation}
\dot{m}_{\rm H}=\frac{\sqrt{2\pi}h_{0}^{2}\Sigma_{0}r_{0}^{2}\Omega_{0}}{1+0.04q_{\rm th}^{2}h_{0}/\alpha}\left(\frac{q_{\rm th}}{3}\right)^{2/3},
\label{eq:mpdotH}
\end{equation}
where $q_{\rm th}=q/h_{0}^{3}$ is the planet thermal mass. 
We can see that the simulated planetary accretion rates can be well reproduced by the theory, 
which is mainly controlled by the disk supply rates.

The lower panel of Figure~\ref{fig:adot_alpha} shows that as the viscosity decreases, the $\dot{a}_{\rm p}/a_{\rm p}$ attributed by the gravitational torque become more negative, 
and thus larger than the less variable positive $\dot{a}_{\rm p}/a_{\rm p}$ from the accretion component in magnitude. 
The transition between the outward and inward migration happens around $\alpha\simeq 0.003$. Such a transition is insensitive to whether we have included the gravitational torque from the sink cell region, but may have a slight dependence on the accretion prescription as discussed above, or other disk parameters which is beyond the scope of this work.

We also show the typical migration rate for non-accreting disks \citep{Kanagawa2018} in Figure~\ref{fig:adot_alpha}. 
Depending on the disk viscosity, the gap depletion effect on the migration rate has been taken into account. 
It can be seen that for $\alpha\lesssim3\times10^{-4}$ the inward migration rate of accreting planets can rougly match to the non-accreting case.  
For a disk with a very low viscosity, the type I migration rate of non-accreting planets is expected to significantly suppressed due to density feedback effect \citep{Li2009,Yu2010}. But it is still unclear the behavior of accreting planets embedded in a much lower viscosity disk. 
There is a tendency that the migration speed would increase towards the lower viscosity as shown in Figure~\ref{fig:adot_alpha} as the positive torque from the CPD disappears and the inner positive Lindblad torque weakens compared to the outer negative Lindblad component due to the disk depletion.  
We should, however, also expect similar feedback effect to occur in very low viscosity disks although the planet mass explored here is much higher than the typical type I regime. 
In the low-viscosity limit, vortices \citep[e.g.,][]{koller2003, likoller2005, Li2020} and/or disk eccentricity \citep[e.g.,][]{Li2021, Dempsey2021} could be also be excited, 
although we do not observe significant eccentricity excitation in our $\alpha=3\times10^{-4}$ 2D run where $K^\prime\equiv q^2/\alpha h_{0}^{3}\simeq27$ is slightly larger than the critical value of $K^\prime\sim20$ identified in 2D non-accreting planet simulations \citep{Dempsey2021}.
They could also alter the migration dynamics of the planet, 
although it could be harder for the disk eccentricity to  be excited in 3D simulations \citep{Li2023}. 
All of these need further numerical simulations to explore in details.

\section{Summary and Discussions}\label{sec:conc}

We have performed 3D/2D hydrodynamical simulations to study the migration for accreting giant planets. A steady mass flux supply from the outer boundary is carefully chosen to ensure that the planetary accretion rate is capped by the supply rate, 
and the inner protostellar disk is accreting steadily onto the central star. 
We find that an accreting planet embedded in a highly viscous disk tends to migrate outward, while it will recover to inward migration for a low viscosity disk, albeit with suppressed rates.
A few 3D/2D simulations have been carried out to confirm the outward migration of the accreting planets, which find overall consistent results. 
The outward migration for the giant accreting planet in a highly viscous disk is due to the asymmetric spiral arm connecting with the CPD region.

By performing a parameter study with the different disk viscosities $\alpha$, 
we find that the transition from outward to inward migration happens around $\alpha\sim0.003$. The accretion rates onto the planet for different viscosities are well predicated by our theoretical formula (Equation~\ref{eq:mpdot_th}).
The circumbinary disk simulations found the inward migration for $q\gtrsim0.01$ with $\alpha\gtrsim0.03$ \citep{Duffell2020b}. 
This dichotomy suggests that, 
for our high viscosity case, there may be another transition from outward migration to inward migration when the mass ratio increases from $0.001$ to $0.01$, which warrants dedicated hydrodynamical simulations in the future.

After the completion of this work, we notice a very similar work by \citet{Laune2024}, who carried out a few 2D hydrodynamical simulations to study the migration of accreting planets and black holes in disks. With slightly different model parameters, e.g., $q=0.001$, $\alpha=0.1$ $h_{0}=0.07$, they also reach the same conclusions as ours, i.e., the accreting object embedded in a highly viscous disk tends to migrate outward with a similar migration rate, although they do not explore the dependence on the disk viscosity.

We have not explored the dependence on other model parameters, such as disk scale height, planet mass, disk self-gravity, and/or disk thermodynamics. 
We have set up the planet in a fixed circular orbit, whereas planet orbital inclination and eccentricity could be other important factors that modify the gravitational and accretion torque. 
The freely released planet could also introduce other dynamical torques on the migrating planets, 
which should co-evolve with the accreting planets. 
We will defer all of these in subsequent studies.

Our results could also be applied to the accreting stellar  and/or intermediate mass black holes embedded in AGN disks. As the turbulence level of AGN disks could approach the order of $\sim0.1$ from gravitational instability \citep{Jiang11,Chen2023}, this could potentially alter the dynamics of black holes migration in disks. Since the migration is nearly stalled around $\alpha\simeq0.003$, this suggests a migration trap for these embedded objects in AGN disks if the viscosity in some regions of AGN disks can cover this range. Such a migration trap could facilitate the formation of binary black holes and the subsequent mergers.

%\section*{Acknowledgements}
\acknowledgements

We thank the referee for many helpful suggestions that improve the quality of the paper. We also thank Dong Lai and Rixin Li for helpful exchanges.
This work is supported in part by the Natural Science Foundation of China (grants 12373070, and 12192223), the Natural Science Foundation of Shanghai (grant NO. 23ZR1473700). The calculations have made use of the High Performance Computing
Resource in the Core Facility for Advanced Research Computing
at Shanghai Astronomical Observatory. 
Softwares: \texttt{Athena++} \citep{Stone2020}, \texttt{Numpy} \citep{vanderWalt2011}, \texttt{Scipy} \citep{Virtanen2020}, \texttt{Matplotlib} \citep{Hunter2007}.

\begin{appendix}

\section{Angular Momentum Flux in the Global Disk} \label{app:amf}

\begin{figure}
\centering
\includegraphics[width=0.45\textwidth,clip=true]{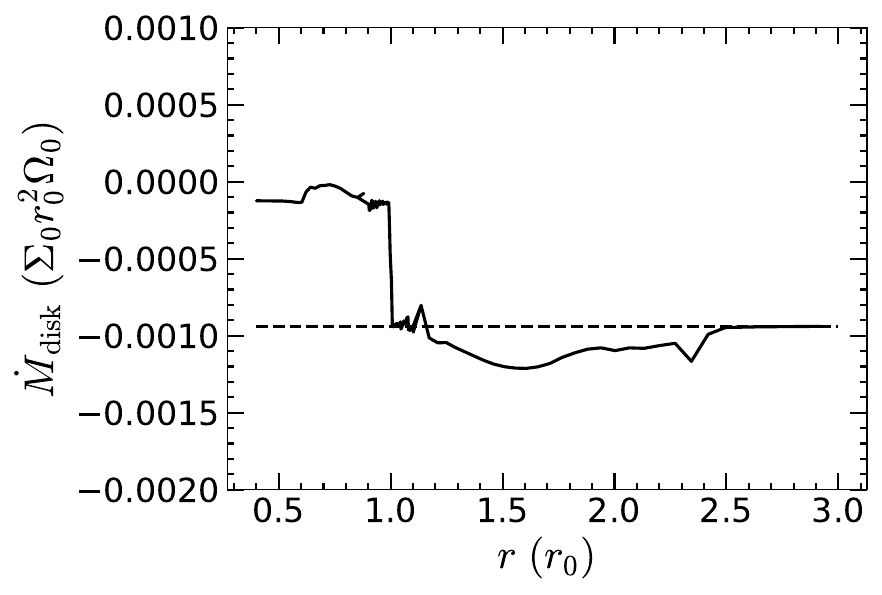}
\includegraphics[width=0.45\textwidth,clip=true]{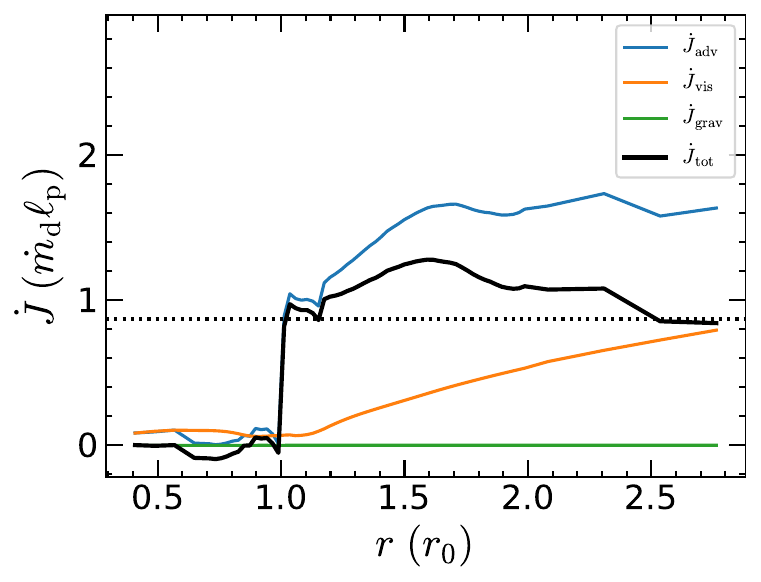}
\caption{The radial profiles of mass flux (upper panel) and angular momentum flux (lower panel) from different components for our 2D run with $\alpha=0.04$ at 10000 orbits (model \texttt{A1}). The dashed line in the upper panel shows the disk supply rate at the outer boundary. The dotted line is the inferred $\dot{J}_{\rm tot}$ onto the accreting planet. The nearly flat profile of $\dot{M}_{\rm disk}$ and $\dot{J}_{\rm tot}$ beyond the planetary orbit, combined with that the jump of $\dot{M}_{\rm disk}$ and $\dot{J}_{\rm tot}$ around the planetary orbit also match the measured $\dot{m}_{\rm p}$ and $\dot{a}_{\rm p}/a_{\rm p}$, suggests the quasi-steady state of the disk.}
 \label{fig:jdot_higv_2d}
\end{figure}

To check the stead state of the disk and the angular momentum exchange in the global disk, we plot the mass flux $\dot{M}_{\rm disk}=\oint r \Sigma v_{r} {\rm d}\phi $ and angular momentum flux $\dot{J}$ in the global disk for a typical 2D model with $\alpha=0.04$ (model \texttt{A1}) in Figure~\ref{fig:jdot_higv_2d}.
The mass flux magnitude remains nearly constant beyond the planetary orbit and then decreases to another constant within its orbit. The mass flux jump at the planetary orbit exactly matches the mass accretion rate onto the planet.
This is an indication of quasi-steady state of the global disk.

The radial angular momentum flux of the disk are composed of different components \citep{Miranda2017}.
In our 2D simulations, the advecting angular momentum flux $\dot{J}_{\rm adv}$ is

\begin{equation}
\dot{J}_{\rm adv}=\oint (-r^{2}v_{r}\Sigma v_{\phi}){\rm d}\phi,
\label{eq:jadv}
\end{equation}
where both $v_{\phi}$ and $v_{r}$ are the quantities measured in the inertial frame. The outward viscous angular momentum transfer rate $\dot{J}_{\rm vis}$ is

\begin{equation}
\dot{J}_{\rm vis}=\oint -r^{3}\nu\Sigma  \left[\frac{\partial}{\partial r} \left(\frac{v_{\phi}}{r}\right) + \frac{1}{r^2}\frac{\partial v_{r}}{\partial \phi}\right]{\rm d}\phi,
\label{eq:jvis}
\end{equation}
and the torque $\dot{J}_{\rm grav}$ from the planet acting on the gas exterior to radius $r$ is

\begin{equation}
\dot{J}_{\rm grav} = \int_{r}^{r_{\rm out}}\oint \left(-r\Sigma \frac{\partial\Phi_{\rm p}}{\partial \phi}\right){\rm d} \phi {\rm d}r.
\label{eq:jgrav}
\end{equation}
The net inward angular momentum flux is thus 

\begin{equation}
\dot{J}_{\rm tot}=\dot{J}_{\rm adv}-\dot{J}_{\rm vis} - \dot{J}_{\rm grav}.
\label{eq:jtot}
\end{equation}
The angular momentum flux for the 3D runs can be calculated similarly by further integrating over $z$.

The different components of these angular momentum flux for model \texttt{A1} are shown in the lower panel of Figure~\ref{fig:jdot_higv_2d}.
The net angular momentum flux $\dot{J}_{\rm tot}$ shows a nearly flat radial profile beyond the planetary orbital radius $r$, and then jumps to zero in the inner disk.
The accretion ``eigenvalue" based on the net angular momentum flux onto the planet is $\ell_{0}\equiv \dot{J}_{\rm tot}/\dot{m}_{\rm p}\simeq1.035\ell_{\rm p}$, as shown with the dotted line in the lower panel of Figure~\ref{fig:jdot_higv_2d}. 

The angular momentum transfer onto the planet $\dot{J}_{\rm p}$, where $J_{\rm p}=m_{\rm p}\sqrt{GM_{*}a_{\rm p}}$, can be obtained from the jump of $\dot{J}_{\rm tot}$ at the planetary orbit.
We can then further calculate the $\dot{a}_{\rm p}/a_{\rm p}$ using the accreted angular momentum flux,

\begin{equation}
\frac{\dot{a}_{\rm p}}{a_{\rm p}}= 2\frac{\dot{m}_{\rm p}}{m_{\rm p}}\left(\frac{\ell_{0}}{\ell_{\rm p}}-1\right) -\frac{\dot{M}_{*}}{M_{*}}.
\label{eq:adot_jtot}
\end{equation}
%We can see that this closely matches the jump of the angular momentum flux around the planet location. 
This leads to $\dot{a}_{\rm p}/a_{\rm p}\simeq 0.07\dot{m}_{\rm p}/m_{\rm p} \simeq 60 \dot{m}_{\rm d}/M_{*}$, which is consistent with the measured $\dot{a}_{\rm p}/a_{\rm p}$ based on the torque calculations shown in Figures~\ref{fig:adot} and \ref{fig:adot_alpha}.
All of these again suggest a quasi-steady state of the global disk.

%The angular momentum flux cross the accretor is larger than the critical

\end{appendix}

\bibliography{references.bib}{}
\bibliographystyle{aasjournalnolink}

\end{document}